\documentclass[preprint2]{aastex}
\shorttitle{RR Lyrae stars in Phoenix}
\shortauthors{Ordo\~{n}ez et al.}

\begin{document}

\title{The RR Lyrae Variable Population in the Phoenix Dwarf Galaxy}

\author{Antonio J. Ordo\~{n}ez\altaffilmark{1}}
\email{a.ordonez@ufl.edu}

\author{Soung-Chul Yang\altaffilmark{2, 3, +}}
\email{sczoo@kasi.re.kr}

\and

\author{Ata Sarajedini\altaffilmark{1}}
\email{ata@astro.ufl.edu}

\altaffiltext{1}{Department of Astronomy, University of Florida, 211
  Bryant Space Science Center, Gainesville, FL 32611, USA}
\altaffiltext{2}{Korea Astronomy and Space Science Institute (KASI), Daejeon, 305-348, South Korea}
\altaffiltext{3}{The Observatories of the Carnegie Institution for Science, 813 Santa Barbara Street, Pasadena, CA 91101, USA}
\altaffiltext{+}{KASI-Carnegie Fellow}

\begin{abstract}
We present the first detailed study of the RR Lyrae variable population in the
Local Group dSph/dIrr transition galaxy, Phoenix, using previously obtained
HST/WFPC2 observations of the galaxy. We utilize template
light curve fitting routines to obtain best fit light curves for RR
Lyrae variables in Phoenix. Our technique has identified 78
highly probable RR Lyrae stars (54 ab-type; 24 c-type) with about 40
additional candidates. We find mean periods for the two
populations of $\langle P_{ab}\rangle = 0.60 \pm 0.03$ days and
$\langle P_{c}\rangle = 0.353 \pm 0.002$ days. We use the properties of these light
curves to extract, among other things, a metallicity distribution
function for ab-type RR Lyrae. Our analysis yields a mean metallicity 
of $\langle [Fe/H]\rangle =
-1.68 \pm 0.06$ dex for the RRab stars. From the mean period and
metallicity calculated from the ab-type RR Lyrae, we conclude that
Phoenix is more likely of intermediate Oosterhoff type; however the
morphology of the Bailey diagram for Phoenix RR Lyraes appears similar to that
of an Oosterhoff type I system. Using the RRab stars, we also study
the chemical enrichment law for Phoenix. We find that our metallicity
distribution is reasonably well fitted by a closed-box model. The parameters of this model are compatible
with the findings of \citet{hid09} further supporting the idea that Phoenix
appears to have been chemically enriched as a closed-box-like system during
the early stage of its formation and evolution. 
\end{abstract}


\section{Introduction}
\label{sec:intro}

With a distance modulus of $(m - M)_0 = 23.09 \pm 0.10$ mag corresponding to a distance of
$d = 415 \pm 19$ kpc and a distance from M31 of $600$ kpc
\citep{hid09}, Phoenix presents an opportunity to study the evolution
of a dwarf galaxy without significant perturbations exerted by massive
galaxies, while still being close enough to obtain good sampling of
its stellar population. Since its discovery \citep{sch76},
Phoenix has gone from being classified as a distant globular cluster
to its currently accepted state as a dwarf transition (dSph/dIrr) type
galaxy, which are characterized by recent star formation but lacking any
prominent H II regions \citep{mat98}. This is supported by
observations of an H I region near the galaxy, likely due to gas
expelled from  supernova winds, that appears to be associated with
recent ($\le$ 100 Myr) star formation \citep{yng07}. Previously, \citet{mart99} discovered
two perpendicular, elliptical components in the structure of
Phoenix. The inner ellipse is oriented in the east-west direction and
contains the young stars in the galaxy. The outer ellipse is rotated
$\sim 90^{\circ}$ from the inner and contains no young stars. This
indicates that either star formation has recently occurred exclusively in the center
of this dwarf galaxy, or stars have formed in an envelope that shrinks
over the time due to the natural reduction of the pressure due to the
gas. 

This last hypothesis was recently suggested in a detailed study
of the star formation history (SFH) of Phoenix performed
by \citet{hid09}. Its distance from massive
galaxies, transition type, associated H I region, and the two
perpendicular, elliptical components with distinctly different stellar
populations 
make the SFH of Phoenix particularly interesting. In their work, \citet{hid09} compared 
synthetic color-magnitude diagrams (CMDs) to the observed CMD of 
Phoenix in order to derive the star formation rate (SFR) as a function
of both time and metallicity. The SFH for the entirety of 
Phoenix was not fit well by any one standard chemical evolution model 
(e.g. closed-box, infall, or outflow; see \citet{pag09, pei94} for
model details) indicating a relatively complex
star formation
history over nearly a Hubble time ($\sim 13$ Gyr). However, they
suggest that a closed-box model is compatible with the SFH
 of Phoenix until about 6-7 Gyr ago when it appears to have
 experienced a sudden burst of chemical enrichment. Thus, an independent
measurement of the abundances of stars that formed during this early epoch probing
the galaxy's chemical evolution at that time 
could test the validity of this analysis.

In this work, we study this early chemical evolution using
the RR Lyrae stars present in Phoenix. The RR Lyrae stars are
pulsating horizontal branch (HB) stars in the instability strip. They are
observed to pulsate in three modes. The ab-types (RRab) pulsate in the 
fundamental mode; the c-types (RRc) pulsate in the first overtone,
while the d-types present both the fundamental and first overtone modes of pulsation.
The discovery of RR Lyrae stars in a system indicates the presence of
an old stellar population 
($\gtrsim 10$ Gyr, \citet{smi95}) characteristic of their
low masses ($\approx 0.6 M_{\odot}$). Thus, in analyzing their 
properties one can probe the conditions of the system at these early 
epochs. Extensive studies of RR Lyrae stars have uncovered many
relations between their pulsation properties and useful astrophysical
quantities \citep{san93, fer98, san06, jur96, alc00, mor07, nem13,
  gul05}. Among these, there is a 
relation between periods, amplitudes, and metallicities of RRab stars
\citep{alc00}. In particular, they define a reduced period, $log$ PA
$= log P + 0.15 A_V$, and found that the
iron abundances, [Fe/H], of the Galactic globular clusters M3, M5, and
M15 correlate with this reduced period.
This provides a straightforward method for deriving the
metallicity distribution function (MDF) for the RR Lyrae population in a system.

\citet{gal04} previously investigated the variable star
population within Phoenix. Specifically, they observed the 
coexistence of anomalous and short-period classical Cepheid variables,
 as well as identified a previously undetected population of RR Lyrae
candidates within the galaxy. That study is the first and only
detection of RR Lyrae stars in Phoenix, but due to observational
constraints (relatively high photometric errors compared with RR Lyrae
pulsation amplitudes), it could not provide an analysis of this population. In this work, we present
the first in-depth study of the RR Lyrae population in Phoenix,
increasing the number of highly probable RR Lyrae stars with light curve
properties by a factor of $\sim 20$. We analyze the properties
 of this RR Lyrae population with the goal of shedding light on the
 early evolutionary history of Phoenix.

This paper is organized in the following manner. Section
\ref{sec:obsred} discusses the observations used in this study and how
these data were reduced. Section \ref{sec:var} describes how variable star candidates were selected and 
characterized as well as how the artificial RR Lyrae simulations were performed in 
order to characterize biases inherent in our analysis. In Section \ref{sec:comp}, we
compare our RR Lyrae sample with candidates identified in \citet{gal04}.
In Sections \ref{sec:res} and \ref{sec:disc}, our results are presented
and discussed. Finally, 
Section \ref{sec:conc} summarizes the conclusions drawn from this work.

\section{Observations \& Data Reduction} 
\label{sec:obsred}

\begin{deluxetable}{ccccccc}
\tabletypesize{\scriptsize}
\tablecaption{\small Observation log. \label{tbl-1}}
\tablewidth{0pt}

\tablehead{
\colhead{Target Field}  & \colhead{RA (J2000)} & \colhead{Dec. (J2000)} & 
\colhead{Filters} & \colhead{Data Set}   & \colhead{HJD Range (+2 454 000)} 
}
\startdata
Inner &  01 51 07.09 & -44 26 40.21 & F555W  &    6x1200            & u64j0101-u64j0106  & 919.98345-920.25843\\
     &              &              & F814W  &    8x1200            & u64j0201-u64j0208  & 920.65355-920.86604\\
\\
Outer &  01 51 08.99 & -44 24 03.94 & F555W  & 1x100,2x1100, 8x1200 & u64j0301-u64j030b  & 925.66641-925.95748\\     
     &              &              & F814W  & 1x100,2x1100,10x1200 & u64j0401-u64j040d  & 926.61775-927.01229\\
\enddata
\end{deluxetable}
 The HST/WFPC2 images of the two target fields around Phoenix used in this 
study were retrieved from the Mikulski Archive
 for Space Telescopes (MAST). The original observing campaign 
(PI: A. Aparicio; GO-8706) was intended to study the spatial structure
and the stellar age and metallicity distribution of this dwarf
galaxy. Therefore, it provides deep time series photometry
with fairly good quality for detecting legitimate RR Lyrae
variable candidates. A detailed description of the two data sets is
summarized in Table \ref{tbl-1}. Images were taken in both
the F555W and F814W filters. A total of two fields were observed: one
centered on Phoenix itself, and the other upon the outskirts of the
galaxy $∼2.\arcmin7$ from the centered field. The total observed field
of view with these observations is equal to 11.4 arcmin$^2$ on the
sky. We note that \citet{hid09} utilized data from
a separate observational campaign (PI: G. Smith; GO-6798) in addition to
the data utilized in this work. Crowding led to difficulties matching stars in
our data reduction. Consequently, we were unable to add this time-series photometry
to the first set, rendering them unusable for our RR Lyrae analysis.

We performed point spread function (PSF) photometry on the data
sets using the HSTPhot package (Dolphin 2000). Each science image 
($*FLT$), which is preprocessed through the standard STScI pipeline 
(bias and dark subtracted, and flat fielded), was cleaned by removing
bad pixels, cosmic rays and hot pixels using the utility software
included in HSTPhot. Pre-constructed PSFs for each WFPC2 passband 
were obtained from the TinyTim PSF library \citep{kri11} and used for our PSF
photometry. Aperture corrections were applied to the output magnitudes
via a default setting in HSTPhot which computes the average
difference between the PSF photometry and aperture photometry with a
0.5 arcsec radius. The resultant magnitudes were also corrected for a
loss of charge transfer efficiency (CTE) for each WFPC2 chip as
described in \citet{dol00}. The HSTPhot package
produces the output photometry both in the native WFPC2 VEGAmag system 
as well as the ground-based Johnson-Cousins system using the
calibration recipe provided by \citet{hol95}. We constructed the final 
list of standard VI photometry by
selecting well-photometered stars with ``object type'' equal to 1
(i.e. good star) and high signal-to-noise ratio, S/N $>$ 10.

\begin{figure*}[!ht]
\epsscale{2.0}
\plotone{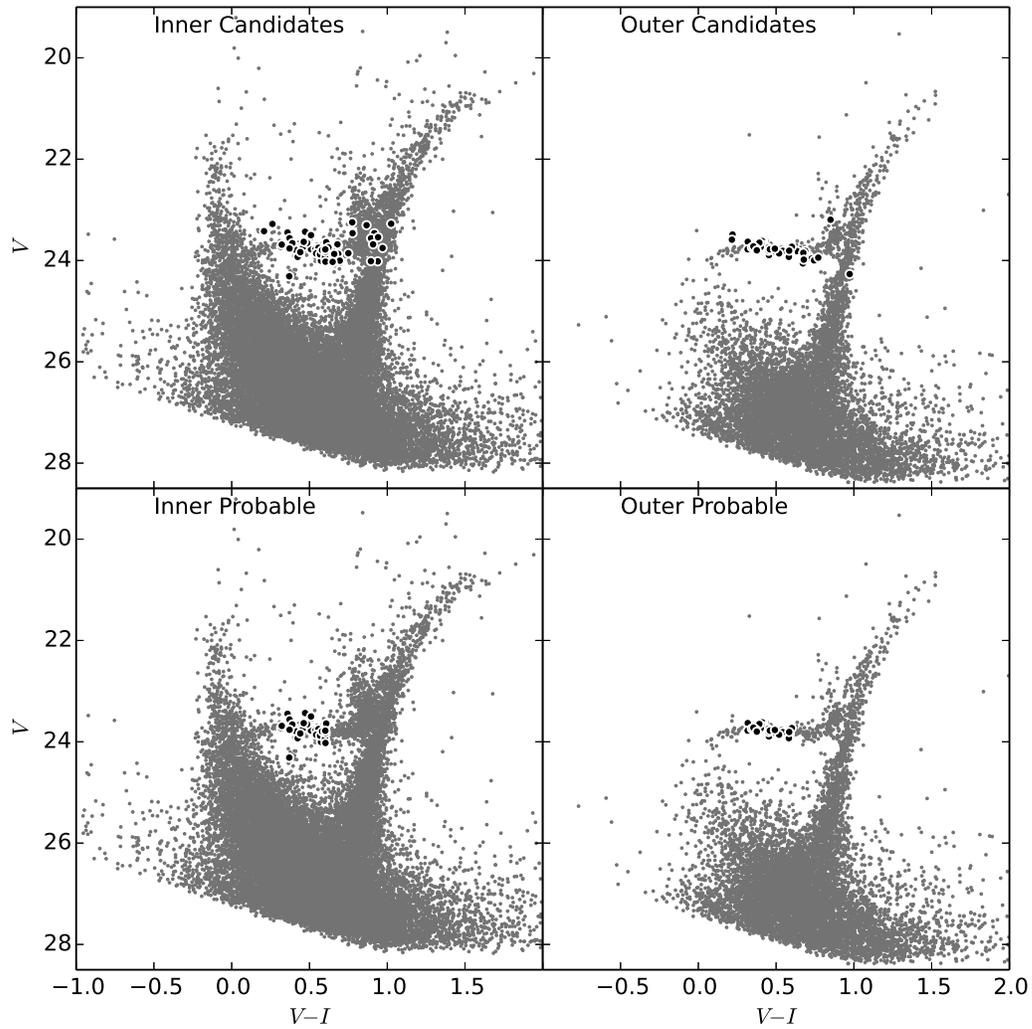}
\caption{{\small The CMDs of Phoenix for each observed field in Phoenix with the
  RR Lyraes marked. The top row shows the CMD with the RR Lyrae
  candidates highlighted, while the bottom row shows the same with
  the probable RR Lyrae stars highlighted.}\label{fig:cmd}}
\end{figure*}

The CMDs for each field observed in Phoenix after our reduction and
photometry are illustrated in Figure \ref{fig:cmd}. \citet{hid09}
present a detailed analysis of the CMD of Phoenix derived from the
same data. Consequently, we only discuss aspects relative to our
analysis. Namely, while only the inner field displays a bright
main-sequence composed of young stars associated with recent star
formation, both the inner and outer fields display a clear red giant
branch and an extended HB indicative of an old
population. The RR Lyrae candidates we detect in our study are marked
on these CMDs. Most lie within the intersection of the HB and the
instability strip with a few outliers. As we will discuss shortly,
most of these outliers are either other types of variable stars or had
poorly sampled light curves. In particular, the inner
field contains more outlying candidates as a consequence of the
fewer observations in this field manifesting in less phase coverage 
in the light curves.

\begin{figure*}[!ht]
\plotone{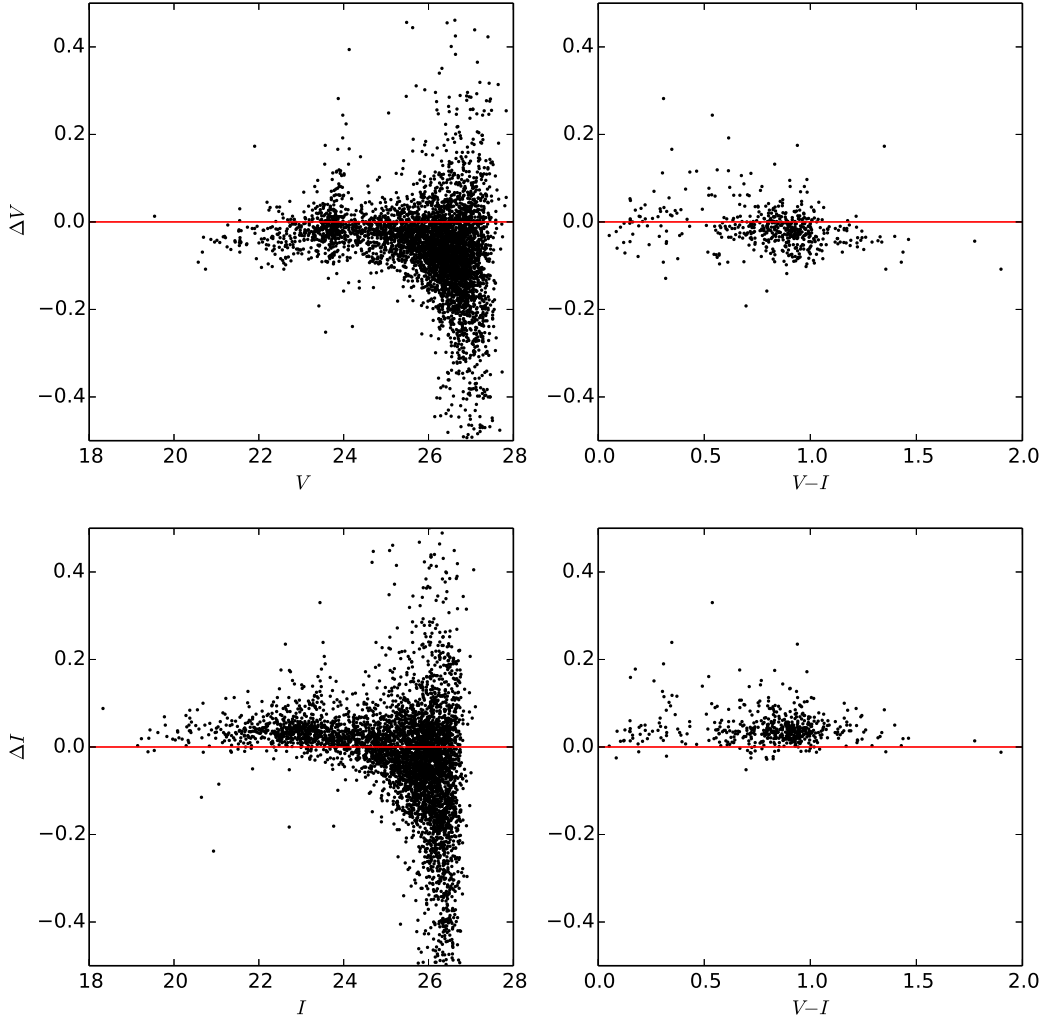}
\caption{{\small Photometry comparison between \citet{hid09} and our
  photometry. \emph{Left:} magnitude difference as a function
  of magnitude for all stars matched between the two photometry
  sets. \emph{Right:} magnitude difference as a function of
  $V-I$ color for matched stars brighter than 24 mag.}\label{fig:photdiffs}}
\end{figure*}

Since \citet{hid09} performed photometry on the same data using the
DAOPHOT package, we looked for differences in the resulting
magnitudes (photometry for comparison kindly provided by S. Hidalgo). 
Figure \ref{fig:photdiffs} illustrates the resulting
differences in photometry for stars matched between our two sets. The
mean offset in $V$ was $\Delta V = -0.015 \pm 0.002$ mag for stars
brighter than $V = 24$ mag, while in $I$ the
offset amounted to $\Delta I = 0.038 \pm 0.002$ mag for stars brighter
than $I = 24$ mag where the reported uncertainties represent the standard
error of the mean. We attribute these small offsets to the
different photometric calibration methods adopted in these two works in
transforming to the Johnson-Cousins system. Considering the small
values of these offsets, we chose not to correct for them and left our
photometry as is for the following analysis.

\section{Variable star characterization and simulations}
\label{sec:var}

We employ the methods of \citet{yng10}, \citet{yng12}, and
\citet{ata12} to characterize the variable stars in the data. 
Stars in a range of V magnitude 
($23$ mag $< V < 24.5$ mag) and $V - I$ color ($-0.2$ mag $< V - I < 1.2$ mag) were
examined for variability. We ranked them by variability using a
reduced $\chi^2_{VI}$ defined as follows:
\begin{equation} \label{eq:chi}
\chi^2_{VI} = \frac{1}{N_V + N_I} \times \left[\sum_{i=1}^{N_V}
  \frac{(V_i - \bar{V})^2}{\sigma_i^2} + \sum_{i=1}^{N_I} \frac{(I_i - \bar{I})^2}{\sigma_i^2} \right]
\end{equation}
where $\bar{V}$ and $\bar{I}$ are the mean magnitudes for each star in
each filter. For each light curve, we ignored data points deviating from the
mean magnitude by more than $\pm 3
\sigma$ from this calculation in order to filter out otherwise stable
stars with anomalous data points. Any star with a $\chi^2_{VI}$ value
greater than $3.0$ was considered a variable star candidate. For
reference the $\chi^2_{VI} $ values of typical non-variable stars at the
level of HB stars of Phoenix [V(HB)$\sim$23.6 mag; see section 4] are less
than 3.0 in our photometry.

\begin{figure}[!ht]
\epsscale{1.0}
\plotone{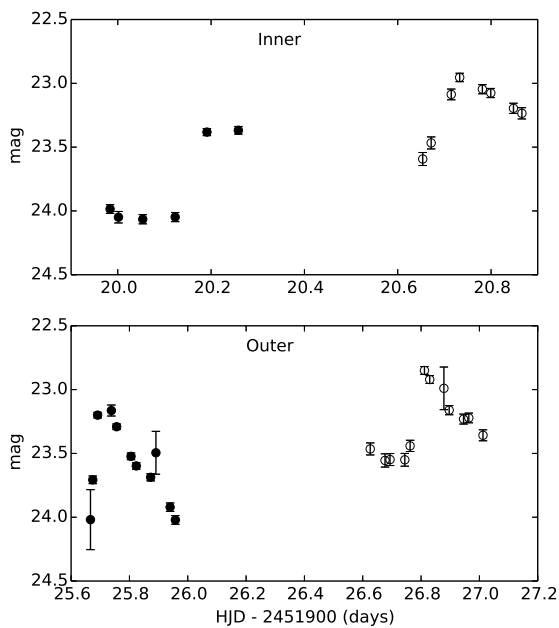}
\caption{{\small \emph{Top:} The raw, unphased light curve for a representative light
  curve in the inner field in Phoenix. Filled points represent
  observations in F555W, while open points are in
  F814W. \emph{Bottom:} Same for the outer field in
  Phoenix.}\label{fig:obswindow}}
\end{figure}

Figure \ref{fig:obswindow} shows the raw, un-phased light curves for two representative
RR Lyrae stars in the inner and outer fields, respectively. The raw
light  curves illustrate qualitatively that our observational window
allows  for adequate detection of variability consistent with RR Lyrae
light curves. Table \ref{tbl:timephot} provides an example time-series
photometry set used for our RR Lyrae fitting routine, and the full set
of time-series photometry for all RR Lyrae candidates is available in
the online version of the journal.

After identifying variable candidates, anomalous data points with high
photometric error ($\sigma \gtrsim 0.1$ mag) were removed from the
time series photometry as these points may act to deform the fitted
light curve from its true shape. We then used FITLC \citep{man08} to
extract best fit RR Lyrae light curves from the 
time series photometry of the candidates. FITLC is a template fitting
program that searches in the RR Lyrae pulsation parameter space
(period, amplitude, mean magnitude) for
the best fitting (minimized $\chi^2$) template. The templates
employed in FITLC are those from \citet{lay00}. FITLC also
provides a GUI that allows the user to visually examine the quality of
the fits as well as their position in $\chi^2$ space.

After this first round of light curve fitting, we implemented a 
constraint on the fit amplitudes according to \citet{df99}. They find
 a relation between the $V$-band and $I$-band amplitudes of RR Lyrae stars, given by
 the following equation:
\begin{equation} \label{eq:ampconst}
A_V = 0.075 + 1.497A_I
\end{equation}
FITLC allows for a constraint on amplitude ratios without a
zero-point offset. For this reason, we use the $A_V$ and $A_I$ amplitude
data from \citet{df99}, which consisted of 127 RR Lyrae
stars from 3 globular clusters,  and performed a linear fit to the
data with zero intercept. This yielded a relation
of the form $A_V = 1.6 A_I$, which agrees with the results of
\citet{liu90}. This ensures that our RRab amplitudes deviate from Equation
\ref{eq:ampconst} by only $\sim 0.01$ mag. The validity
 of these fits was then checked by examining positions in the Bailey 
diagram, the best fit light curves by eye, and adherence to Equation \ref{eq:ampconst}. Stars with anomalous fits 
(e.g. RRab stars with extremely low or high periods due to aliasing) were examined 
and then manually re-fit using the interactive fitting mode in
FITLC. The stars needing human intervention to be properly fit
amounted to $\sim 30$. A total of 121 RR Lyrae candidates were identified with this method. Figure \ref{fig:cmd} shows the 
locations of these candidates in the CMD of Phoenix. 

In order to minimize contamination of our RR Lyrae sample from other
variables, we performed a color cut
excluding stars outside of the instability strip. Utilizing the instability strip bounds from \citet{mg03}
($0.28$ mag $< V-I < 0.59$ mag) and a reddening of $E(V-I)=0.02$ mag from the
recalibrated dust maps of
\citet{sch98} \citep{sf11}, we narrowed our RR 
\begin{figure*}[!ht]
\epsscale{2.0}
\plotone{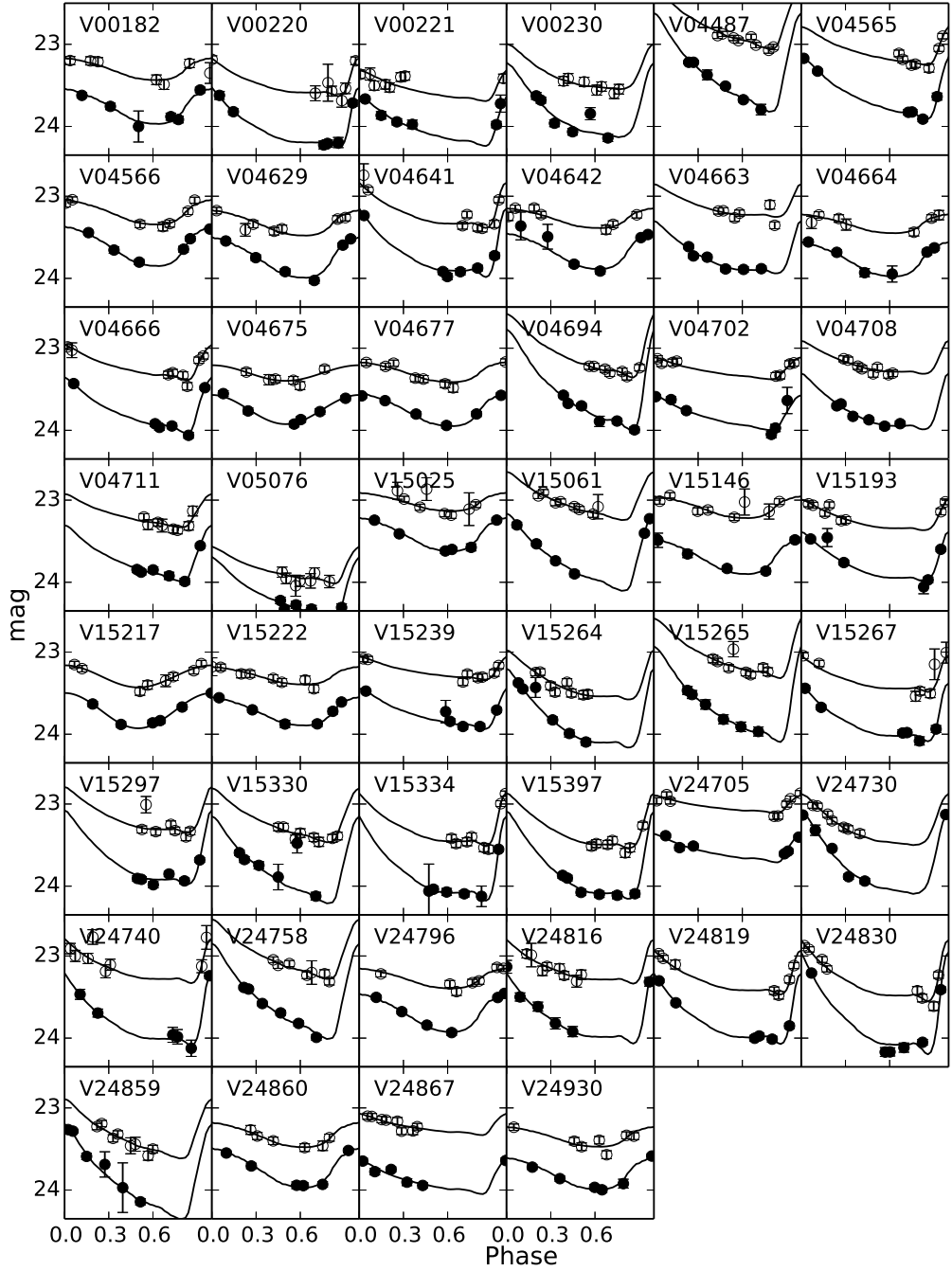}
\caption{{\small The best fitting light curves for the probable RR Lyrae stars in the inner
  Phoenix field. The filled circles represent the $V$-band data while
  the open circles are the $I$-band data.}\label{fig:disklc}}
\end{figure*}

\clearpage

\begin{figure*}[!ht]
\plotone{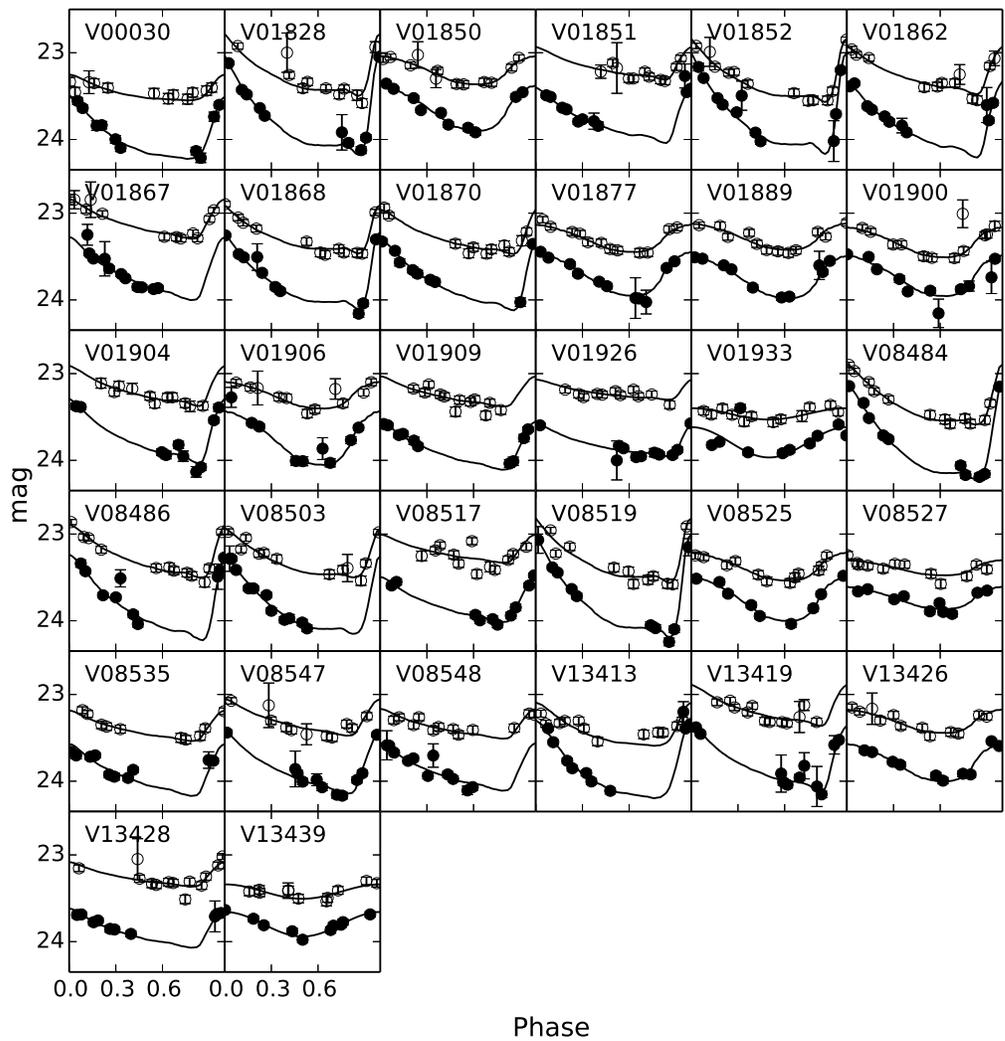}
\caption{{\small Same as Figure \ref{fig:disklc} for the outer field.\label{fig:halolc}}}
\end{figure*}

 \clearpage

\noindent Lyrae list to 78 members
within this region of the CMD. Figures \ref{fig:disklc} and \ref{fig:halolc}
display the fitted light curves for these probable RR Lyrae stars for the inner and
outer fields, respectively. This resulted in two sets of RR Lyrae
candidates within Phoenix. The first set contains the 121 variables
with best fitting
periods, amplitudes, and light curves appropriate for RR Lyrae stars
which we will refer to as the RR Lyrae candidates in Phoenix, and
the second set consists of the 78 members of the first set that are located
within the color range of the instability strip, which we will henceforth
refer to as the probable RR Lyrae stars in Phoenix. Tables
\ref{tbl:rrlconf} and \ref{tbl:rrlcan} list the light curve fit parameters
for the probable and the candidate RR Lyrae stars, respectively. In Section
\ref{sec:res}, we analyze and compare the properties of both sets.
To characterize and quantify biases inherent in our fitting
procedure, we produced artificial RR Lyrae light curves and used FITLC
to retrieve the best fit properties of these artificial variables
given our observing window. We generated 500 synthetic light curves
per RR Lyrae template (6 RRab; 2 RRc) and simulated observations of these
variables with the window function of our data set. These simulated
observations were used as input to FITLC from which output fit
parameters were generated. The deviations of the fit parameters from
the true light curve parameters provide a quantitative estimate of the
biases resultant from the phase coverage and cadence of the data.

\begin{figure}[!ht]
\epsscale{1.15}
\centering
\plotone{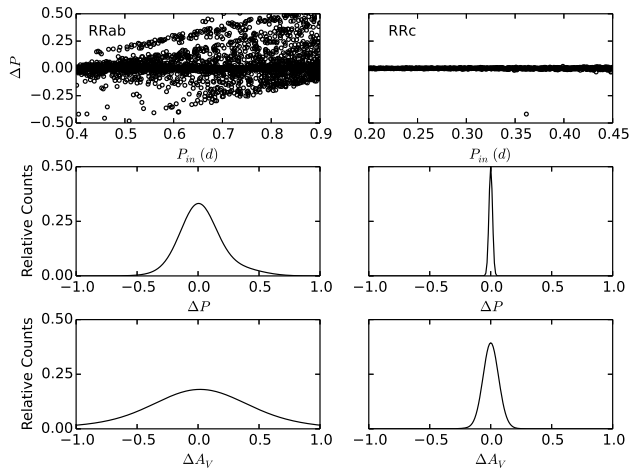}
\caption{{\small Results from the artificial RR Lyrae simulations for an observing
  cadence matching the inner Phoenix field.}\label{fig:disksim}}
\end{figure}

\begin{figure}[!ht]
\plotone{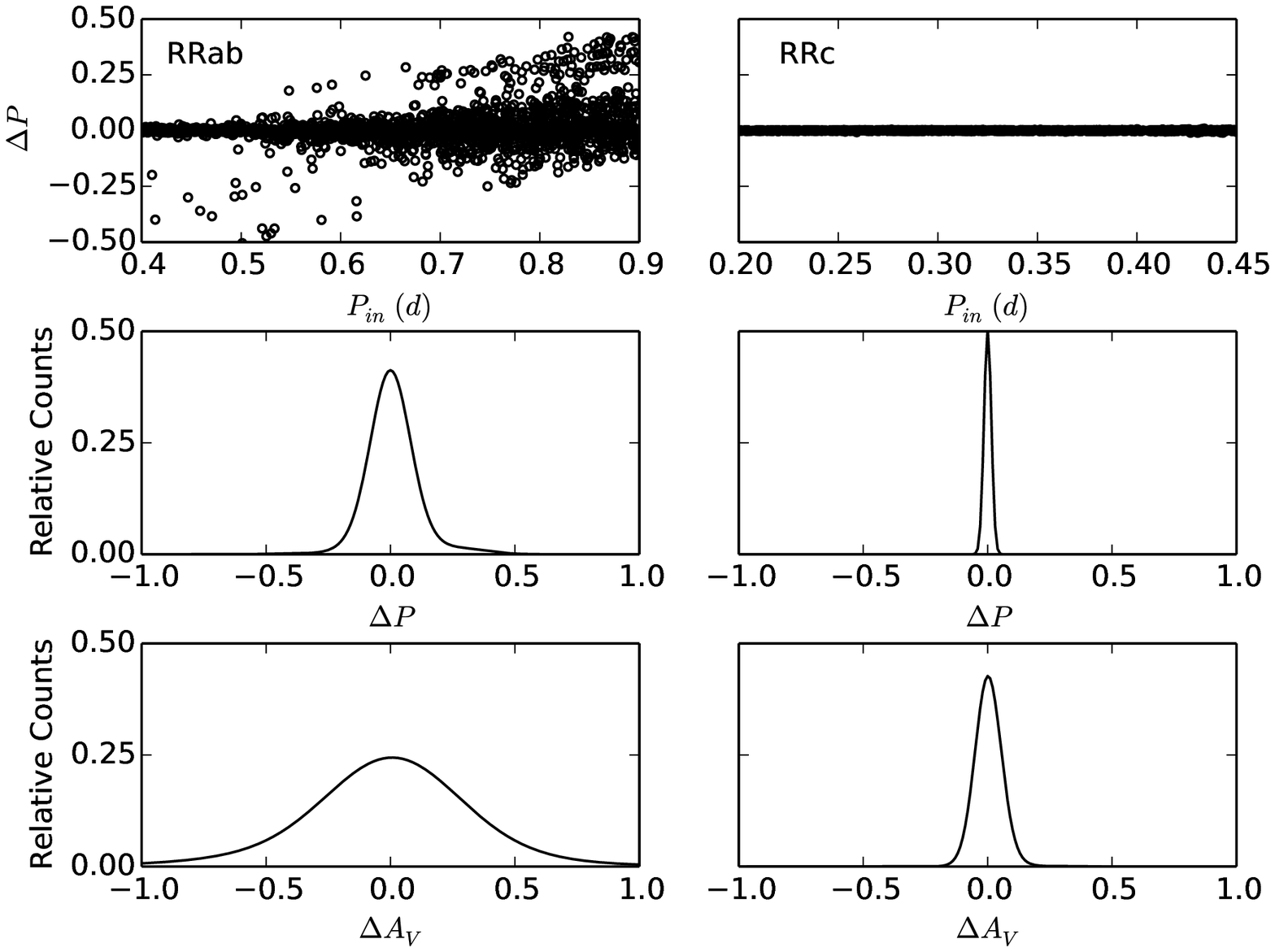}
\caption{{\small Results from the artificial RR Lyrae simulations for an observing
  cadence matching the outer Phoenix field.}\label{fig:halosim}}
\end{figure}

Figures \ref{fig:disksim} and \ref{fig:halosim} illustrate the
results of these simulations for both of the observed fields in Phoenix. It
is evident that while the outer field observations suffer from relatively
weak aliasing effects, the inner field shows significant alias
bands resulting from the shorter phase coverage of those
observations. In order to quantify the uncertainties present in the fitted RR Lyrae
properties due to these alias bands,
we performed the statistical test used in \citet{yng10}, \citet{yng12},
and \citet{ata12} with a modification. This consists of sampling the set of artificial RR
Lyrae stars 
and randomly drawing the appropriate number corresponding to the
number of probable RR Lyrae stars observed in each field. For the inner field, 32 RRab
and 14 RRc stars were randomly sampled from the artificial RR Lyrae
stars with simulated observations corresponding to that of the inner
field data set. Likewise, for the
outer field, 22 RRab and 10 RRc stars were
sampled from the corresponding set of artificial RR Lyrae stars in the
outer field. From these samples, the mean deviations
of the fitted parameters from the true parameters ($\langle \Delta P\rangle$, 
$\langle \Delta A_V\rangle$, and $\langle \Delta (V-I)_{min}\rangle$) were calculated for each type in each field. This
process was repeated 10,000 times to increase the statistical
significance. 

\begin{deluxetable}{cccccc}
\tabletypesize{\scriptsize}
\tablecaption{\small Results from the artificial RR Lyrae
  simulations.\label{tbl:err}}
\tablewidth{0pt}
\tablehead{
\colhead{Field} & \colhead{RR Lyrae type} & \colhead{Total period
  error (days)} & \colhead{Total amplitude error
(mag)} & \colhead{Total $(V-I)_{0, min}$ error (mag)} & \colhead{Total
$\phi_{31}$ error }
}
\startdata
Inner & RRab & $\pm 0.0312$ & $\pm 0.0118$ & $\pm 0.0144$ &$\pm  0.0132$ \\
Inner & RRc & $\pm 0.0003$ & $\pm 0.0033$ & N/A & N/A \\
Outer & RRab & $\pm 0.0071$ & $\pm 0.0128$ & $\pm 0.0068$ & $\pm 0.0186$ \\
Outer & RRc & $\pm 0.0024$ & $\pm 0.0053$ & N/A & N/A \\
\enddata
\end{deluxetable}

We then estimate the uncertainties in these fit properties from the
distributions of the deviations as follows. We first fit a Gaussian to
each distribution. The 1-$\sigma$ width of this Gaussian approximates
the standard deviation, thus the standard error of the mean (sem) ($\sigma / \sqrt{N}$)
 approximates the random error associated with this bias. Meanwhile the 
peak, $\mu$, is a measure of the systematic
bias resulting from alias bands not symmetric about zero. Herein lies
 the modification of this error determination technique from the
 previous studies that employed it, as they did not account for this
 systematic offset. We then add
these distinct components in quadrature to realistically estimate the
total error inherent
in the fitting process as $s = \sqrt{(\sigma / \sqrt{N})^2 +
  \mu^2}$. This represents the total standard error inherent in the fitting procedure
for each quantity ($P$, $A_V$, $(V-I)_{min}$), type (RRab; RRc),
and field (inner; outer). The resulting errors are tabulated in Table
\ref{tbl:err}. These errors indicate that our fit parameters
are sufficiently accurate and precise for the analysis to follow. We
conclude this section by noting that these estimations
represent the upper limits on our actual uncertainties since they
consider only the automated fitting routine. As was pointed
out earlier in this section, approximately one third of RR Lyrae stars
in our sample were fit with human intervention which suffers from lower uncertainty.

\section{Comparison with previous study}
\label{sec:comp}

\begin{figure}[!ht]
\epsscale{1.25}
\plotone{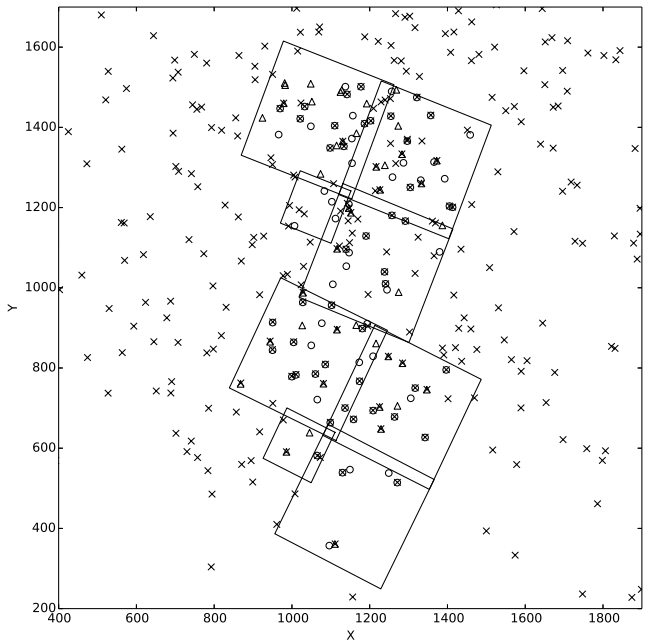}
\caption{{\small Positions of RR Lyrae candidates detected in Phoenix. X and Y are
  arbitrary coordinates on the sky. The crosses
  are candidates from Gallart et al. (2004) while the open circles are the
  probable RR Lyrae stars found in this work. The
  positions of the RR Lyrae candidates outside of the instability
  strip are also plotted as open triangles. The boundaries of the WFPC2 chips are
  shown for reference.}\label{fig:rrlpos}}
\end{figure}
In this section, we compare the properties of the RR Lyrae stars found in
this study to those of \citet{gal04}. The data for this comparison were
kindly provided by C. Gallart (private communication). We first
compared the locations on the sky of the RR Lyrae candidates from the two sets. Figure
\ref{fig:rrlpos} shows the positions of the RR Lyrae candidates detected in this
work (open circles) superimposed on the positions of the $\sim$ 400 RR Lyrae candidates from
the previous study (crosses). It is clear that we identify many  RR Lyrae
candidates in common,
although the previous study detected many at locations where our analysis
does not and vice versa. Since the horizontal branch of Phoenix was near the photometric limit
of the \citet{gal04} work, the pulsation properties of their RR Lyrae
candidates remain largely uncertain. Consequently,
the authors could only confirm a fraction of these RR Lyrae stars. They
calculated periods for $4$ of these, all of which have high
amplitudes ($A_V \approx 1$ mag). Thus, it is  very  likely that some
candidates from their sample are false-positives since pulsation
amplitudes smaller than $A_V \lesssim 1$ mag approach the photometric error
of their observations.

\citet{gal04}
provide light curve fit properties for four of these RR Lyrae stars,
of which three lie in our field of view. We find reasonable matches in
our sample for two of them. These two
are the RRab candidates $V01920$ (outer field) and $V04721$ (inner
field), labeled as $4439$
and $7008$ in \citet{gal04}.  Our fitting routine identified these as
RRab stars with periods of $P = 0.6058$ d and $P = 0.6139$ d, and
mean $V$-band magnitudes of $\langle V\rangle = 23.63$ mag and
$\langle V\rangle = 23.70$ mag. In
comparison, \citet{gal04}, using similar template light curve fitting
routines, obtained periods of $P = 0.7580$ d and $P = 0.6317$
d, and mean magnitudes of $\langle V\rangle = 23.56$ mag and $\langle
V\rangle = 23.61$ mag, respectively. Their
periods deviate from our calculated periods by
$\Delta P_{V01920} = 0.1522$ d and $\Delta P_{V04721} = 0.0178$
d, while our mean magnitudes differ by $\Delta \langle V\rangle_{V01920} =
-0.07$ mag and $\Delta \langle V\rangle_{V04721} = -0.09$ mag. 

Although
quantitative comparison is difficult with only two data points, we
note that the differences in mean magnitudes are small and within
the photometric errors. The period deviation for $V04721$ is small and
within the RRab period error for the inner field listed in Table
\ref{tbl:err}. However, the period deviation for $V01920$ is higher
than our calculated RRab period error for the outer field by a factor
of $\sim 20$. Although \citet{gal04} did not provide uncertainties for
their RR Lyrae periods, it may be possible that this was simply an
erroneous fit due to the high photometric errors
involved.

We conclude this comparison by emphasizing that our most conservative
set of 78 RR Lyrae stars increases the number of probable RR Lyrae
variables with light curve parameters calculated in
Phoenix by a factor of $\sim 20$. Additionally, the range of RR Lyrae amplitudes
we detect span the full range of pulsation amplitudes
allowing for a good sampling of period-amplitude space.

\section{Results}
\label{sec:res}

We now turn to the results of our analysis of the RR
Lyrae population in Phoenix. We first examine the Bailey diagram for
these RR Lyrae stars and discuss their Oosterhoff classification. 
We then calculate the line of sight reddening to Phoenix
using the minimum light colors of the RRab light curve
fits. Finally, we use an empirical relation between the period and
amplitude of RRab stars and their metallicities to study the early
chemical evolution of Phoenix as manifested in its RR Lyrae stars.

\subsection{Bailey Diagram}
\label{sub:bailey}

\begin{figure*}[!ht]
\epsscale{2.0}
\plotone{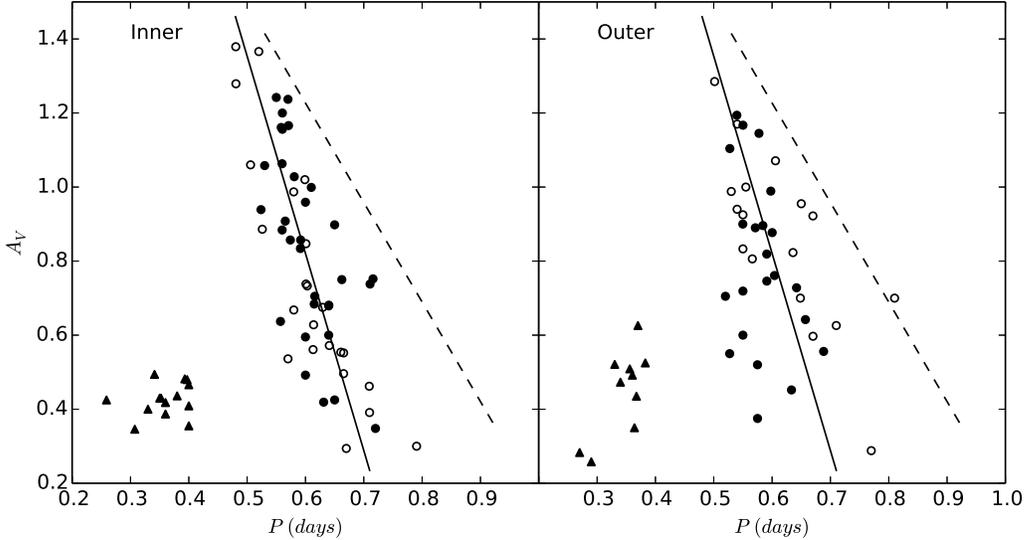}
\caption{{\small The Bailey diagrams for the RR Lyrae populations in the inner
  (left) and outer (right) fields observed within Phoenix. The
  circles are RRab types while the triangles are RRc types. Filled
  points correspond to probable RR Lyrae stars, while open points are
  candidates outside of the instability strip. The
  loci of the RRab types Oosterhoff type I and II Galactic globular
  clusters \citep{cr00} are also plotted as the solid and dashed line, respectively.}\label{fig:bailey}}
\end{figure*}
Figure \ref{fig:bailey} shows the Bailey diagrams from each observed field for the RR Lyrae stars we
detected in Phoenix. The appearance of these diagrams does not indicate
any significant difference in RR Lyrae properties between the two
fields. The average periods for each type in each field
are tabulated in Table \ref{tbl:meanp}. The difference between the average RRab periods in
the two fields, $\Delta \langle P_{ab}\rangle = 0.021$ d, is
comparable to the period errors for RRab stars derived in Section
\ref{sec:var}, thus we regard this difference as statistically
insignificant. However, the difference between the average RRc periods
in the two fields, $\Delta \langle P_c\rangle = 0.016$ d, while small, is
significantly greater than the derived period errors for RRc stars
listed in Table \ref{tbl:err}. We could not identify any apparent
reason for this discrepancy but note that our analysis primarily
utilizes the properties of the RRab stars. Thus, we do not anticipate
this effect to influence our subsequent analysis.
\begin{deluxetable}{ccc}
\tablecaption{\small The mean periods of each RR Lyrae type found in each
  observed field.\label{tbl:meanp}}
\tablewidth{0pt}
\tablehead{
\colhead{Set} & \colhead{$\langle P_{ab}\rangle$ (d)} &
\colhead{$\langle P_c\rangle$ (d)}
}
\startdata
Inner Candidates & 0.605 & 0.359\\
Outer Candidates & 0.598 & 0.343\\
Inner Probable & 0.603 & 0.359\\
Outer Probable & 0.582 & 0.343\\
\enddata
\end{deluxetable}
Combining both fields, we find mean periods of $\langle P_{ab}\rangle = 0.595 \pm
0.007$ d (sem) and $\langle P_c\rangle = 0.353 \pm 0.008$ d (sem) for the two RR Lyrae
types residing within the instability strip in Phoenix. When we
include the RR Lyrae candidates outside of the instability strip, the
mean period of RRab stars shifts to a slightly higher value of
$\langle P_{ab}\rangle = 0.602
\pm 0.007$ d, while $\langle P_c\rangle$ remains unchanged since no c-types were
found outside of the instability strip.  

The loci of the RRab types in Oosterhoff
type I (OoI) and II (OoII) Galactic globular clusters from \citet{cr00} are also plotted in Figure
\ref{fig:bailey}. The ab-types in Phoenix appear to follow the OoI
relation quite well, with the exception of one RRab candidate that appears near
the OoII locus. This candidate is $\sim 1$ magnitude brighter than
most other RR Lyrae stars in our sample and is likely an anomalous
Cepheid contaminating our sample.

According to \citet{cat09}, the Oosterhoff dichotomy describes the
tendency of Galactic halo globular clusters to be divided into the two
Oosterhoff types with mean RRab periods of $\langle P_{ab}\rangle \approx 0.55$ d for OoI
and $\langle P_{ab}\rangle \approx 0.65$ d for OoII. In contrast, dSph galaxies
have been observed to be of intermediate Oosterhoff type, with $0.58
\le \langle P_{ab}\rangle(d) \le 0.62$, referred to as the Oosterhoff gap. The primary factor influencing the Oosterhoff type of a
system is thought to be the metallicity of that stellar population,
although other factors such as age are believed to be partially responsible for this dichotomy as
well \citep{lee99}. OoI systems are generally of intermediate
metallicity while OoII systems tend to be more metal-poor \citep{lee99, cat09}. Clearly,
the mean period we calculate for probable RRab stars in Phoenix of
$\langle P_{ab}\rangle = 0.595$ d places it well within the Oosterhoff
gap, albeit slightly closer to OoI systems. We will show momentarily that the uncertainty in mean period as
calculated from Section \ref{sec:var} admits a significant possibility of Phoenix
being an OoI system as opposed to OoII.

\begin{figure}[!ht]
\epsscale{1.1}
\plotone{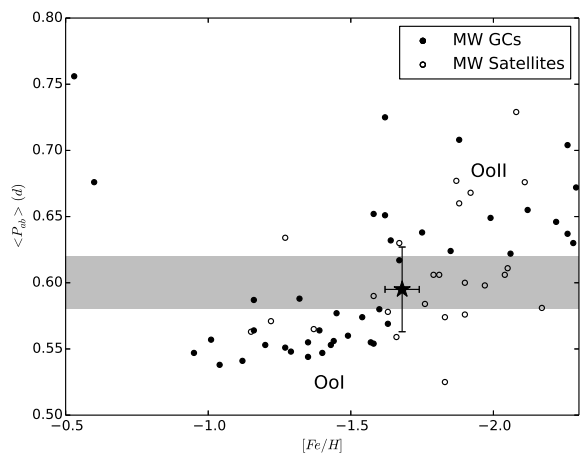}
\caption{{\small [Fe/H] vs $\langle P_{ab}\rangle$ for Galactic globular clusters and
  Milky Way dwarf satellites from \citet{cat09}. The location of
  Phoenix in this diagram is marked with a star. In this context,
  Phoenix appears as an Oo intermediate system, accompanying most
  other dwarf galaxies orbiting the Milky Way.}\label{fig:pmet}}
\end{figure}

Figure
\ref{fig:pmet} illustrates the Oosterhoff dichotomy in terms of the
metallicity of a system and $\langle P_{ab}\rangle$ of its RR Lyrae population
\citep{cat09}. In contrast to the appearance of the Bailey diagram
of Phoenix, its intermediate metallicity (calculated from the RRab
stars; see Section \ref{sub:met}) and $\langle P_{ab}\rangle$ place
it well  within the Oosterhoff gap, amongst most other
Milky Way satellite systems. Adding to this, Phoenix appears to
contain a higher fraction of longer period ($P_{ab} \gtrsim 0.6$ d),
lower amplitude ($A_V \lesssim 0.5$ mag) RRab stars
in comparison to the canonical OoI systems. However, we note that the
error bars for the period of Phoenix in this plot, calculated from the
artificial RR Lyrae simulations in Section \ref{sec:var}, do allow for
the possibility of Phoenix to be an OoI system. We also acknowledge
the possibility that the window function of our data may have
prevented detection of some shorter period RRab stars. While we cannot
rule this possibility out, its position well within the Oosterhoff gap in Figure
\ref{fig:pmet} leads us to conclude that Phoenix is more likely of intermediate
Oosterhoff type.

\subsection{The reddening to Phoenix from the RRab stars}
\label{sub:red}

It has been previously determined that the intrinsic ($V-I$) colors of RRab stars at the
minimum phase of their pulsation is restricted to a small
range. \citet{gul05} calculated this minimum light color to be
$(V-I)_{0, min} = 0.58 \pm 0.02$ mag. We use this value to calculate the
reddening to the individual RRab stars in Phoenix and construct a
reddening distribution. 

\begin{figure}[!ht]
\plotone{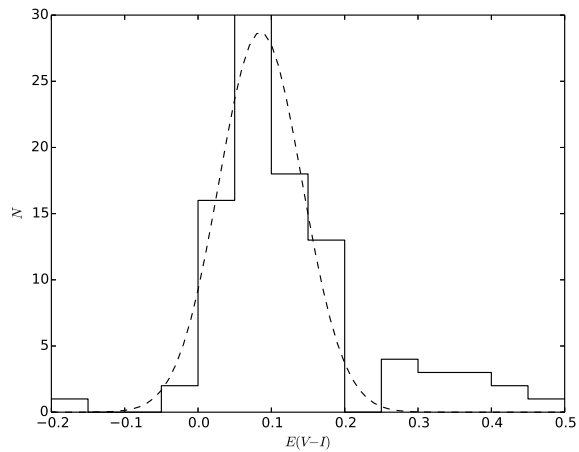}
\caption{{\small The reddening distribution of the RRab candidates in Phoenix. The
  individual reddenings were calculated using the apparent minimum
  light color, $(V-I)_{min}$, and the
  intrinsic minimum light color, $(V-I)_{0, mn}$ from Guldenschuh et al. (2005).}\label{fig:canred}}
\end{figure}

\begin{figure}[!ht]
\plotone{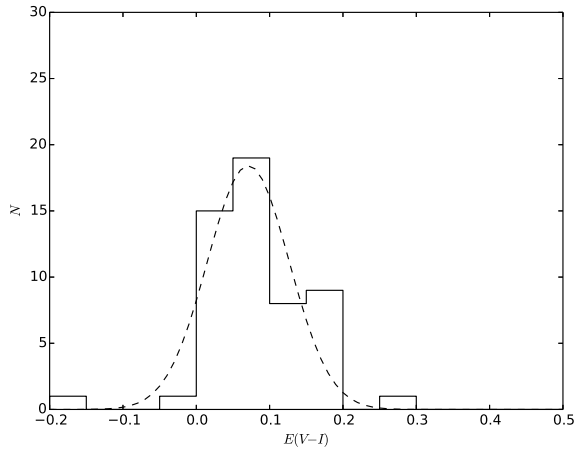}
\caption{{\small Same as Figure \ref{fig:canred} with RRab candidates outside
  of the instability strip excluded.}\label{fig:red}}
\end{figure}

Previous studies of the interstellar medium in Phoenix conclude that
the reddening to stars within the galaxy is fairly small [$E(B-V) = 0.03-0.07$
mag, \citet{bia12}]. Thus, we expect little to no differential reddening within
Phoenix and combine the reddening distributions from both the inner and
outer fields. To calculate the reddening to each individual RRab star,
we used the best fit light curves to obtain the colors at minimum
light, which we then converted to reddenings, $E(V-I)$, using the
relation from \citet{gul05}. The reddening distributions are illustrated in Figure \ref{fig:canred} for the
candidates and Figure \ref{fig:red} for probable RR Lyrae stars in Phoenix. 

To obtain a singular value for the line of sight
reddening and an error estimation, we fit a Gaussian to each
distribution. The peak of this Gaussian approximates the mean
reddening value while the 1-$\sigma$ width approximates the standard
deviation of this
value. We take the standard error of the mean to
estimate the random error of this value. Two additional sources of
error for the reddening values result from the
uncertainty in the intrinsic minimum light color of RRab stars from
\citet{gul05} and the uncertainty in the minimum light
colors we calculated from the artificial RR Lyrae simulations (see
Table \ref{tbl:err}). We add these sources of uncertainty together in
quadrature to obtain a realistic error estimate for this
reddening value to Phoenix. This results in a reddening of
$E(V-I) = 0.09 \pm 0.06$ mag using all the RRab candidates and
$E(V-I) = 0.07 \pm 0.06$ mag utilizing only the probable RRab stars. Additionally, we
repeated this process separately for both observed fields to check our
initial assumption of little to no differential reddening. We obtain
for the probable RRab stars in Phoenix $E(V-I) = 0.05 \pm
0.04$ mag for the inner field and $E(V-I) = 0.08 \pm 0.04$ mag for the outer
field. The difference in reddening between the two fields is within the errors, thus we
regard it as statistically insignificant. 

These reddening
values do not agree well with that of Phoenix from the
  recalibrated \citet{sch98}
maps \citep{sf11} of $E(V-I) = 0.02$ mag. However, previous studies of the stellar
populations in Phoenix have found the $E(B-V)$ reddening to be higher in
Phoenix than from the \citet{sch98} maps. \citet{bia12} found a
reddening between $E(B-V) = 0.03 - 0.07$ mag using SED fitting. Using the
location of the plume of blue supergiants in its CMD, \citet{mas07}
calculated a larger reddening of $E(B-V) = 0.15$ mag. The comparison
  of these values with
the reddening from the recalibrated \citet{sch98} maps of $E(B-V) =
0.01$ mag indicates
a significant source of internal reddening within Phoenix. Taking into
account that E(V-I)=1.62*E(B-V) \citep{car89}, we find that
our results qualitatively agrees with these previous findings.

\subsection{The metallicity of RRab stars in Phoenix}
\label{sub:met}

We use the relation between RRab period, amplitude, and
metallicity from \citet{alc00} to calculate the metallicity of the
RRab stars in Phoenix:
\begin{equation} \label{eq:met}
[Fe/H] = -8.85[log P_{ab} + 0.15A_V] - 2.60
\end{equation}
This yields a mean metallicity for all RRab candidates in Phoenix of
$\langle [Fe/H]\rangle = -1.69 \pm 0.05$ dex. This value is essentially unchanged
when the RRab stars outside of the instability strip are excluded: $\langle [Fe/H]\rangle
= -1.68 \pm 0.06$ dex. The uncertainties quoted for these metal abundances
reflect the standard error of the mean summed in quadrature with the
period and amplitude errors from Table \ref{tbl:err} propagated
through the metallicity calculation. \citet{hel99} previously estimated
the mean metal abundance of Phoenix by comparing the RGB in the $V$,
$(V-I)$ CMD. Through this method, they obtained a mean abundance of
$[Fe/H] = -1.81 \pm 0.10$ dex with which our abundance calculation from
the RR Lyrae population agrees within the uncertainties. Meanwhile,
\citet{hid09} calculated the mean metallicity of Phoenix to be
$\langle [Fe/H]\rangle = -1.7 \pm 0.2$ dex from their SFH analysis (where
they assumed $[Fe/H] = [M/H]$), with which our metallicity also agrees
well.

An alternative method for determining the metallicities of RRab stars
involves using the shape of their light curves. \citet{jur96}
originally calibrated this relation using the period and Fourier
parameter $\phi_{31}$, finding that:
\begin{equation} \label{eq:metjk96}
[Fe/H] = -5.038 - 5.394P_{ab} + 1.345\phi_{31}
\end{equation}
More recently, \citet{nem13} re-calibrated this relation using RR
Lyrae stars in the \emph{Kepler} field, finding:
\begin{equation} \label{eq:metn13}
[Fe/H] = -8.65 - 40.12P_{ab} + 5.96\phi_{31} + 6.27P_{ab}\phi_{31} - 0.72\phi_{31}^2
\end{equation}
We note that it is necessary to first convert the $\phi_{31}$ values
in the $V$-band to the \emph{Kepler} system before computing the metallicities this
way. We use the calibration of \citet{nem11} for this conversion,
where they find that $\phi_{31}(Kp) = \phi_{31}(V) + (0.151 \pm
0.026)$. \citet{alc00} compared their calibration with that of
\citet{jur96}, noting that metallicities using the latter relation
must be shifted by -0.2 dex in order to place them on the same
metallicity scale. Since the \citet{nem13} calibration is on the same
scale as \citet{jur96}, we shift the metallicities using both
relations by -0.2 dex to place everything on the \citet{alc00}
metallicity scale.

A direct Fourier decomposition of the dataset was not considered to
be the most effective method of determining $\phi_{31}$ given the low
number of data points. Instead, we opted to use the method of
template Fourier fitting as described by \citet{kk07}, which is better
suited for light curves containing less than 20 data points. In
particular, we utilized the Template Fourier Fitting (TFF) code provided by \citet{kk07} for the
fitting process using the FITLC determined periods in conjunction with the
$V$-band light curves as input. The Fourier
amplitude and phase coefficients are then calculated by comparing with
248 RRab light curve templates. The phase coefficient, $\phi_{31}$,
was extracted from these TFF fits and used to determine metal
abundances for the probable RRab stars in Phoenix using the two
relations previously mentioned. 

\begin{figure}[!ht]
\epsscale{1.25}
\plotone{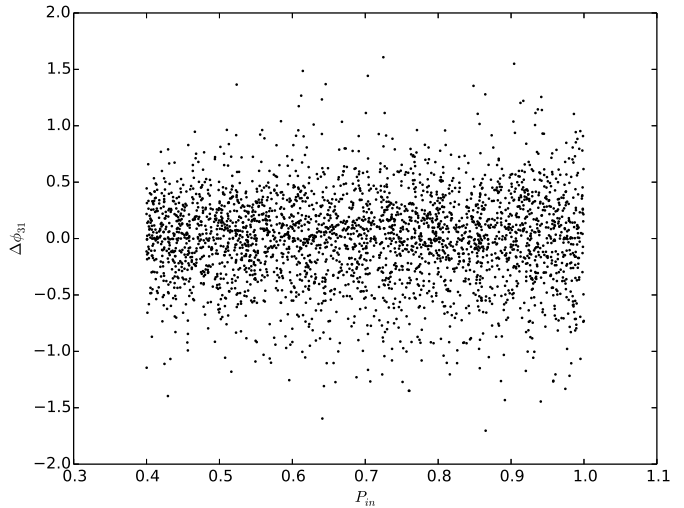}
\caption{{\small Difference between the input and output $\phi_{31}$ versus
  input period from the
  TFF artificial RR Lyrae simulations for the outer field. The simulations show no trends
  in $\phi_{31}$ error with period, indicating no systematic errors
  in $[Fe/H]$ with period for the TFF metallicities. The simulations
  from the inner field also lack any trends with period.}\label{fig:phi31diff}}
\end{figure}

We performed artificial RR Lyrae
simulations similar to those described in Section \ref{sec:var} in
order to estimate the errors in $\phi_{31}$. The only
modification was to run TFF on the artificial RR Lyrae stars using the
periods calculated by FITLC. The errors derived in this way are also
tabulated in Table \ref{tbl:err}. Figure \ref{fig:phi31diff}
illustrates the differences in the input and output $\phi_{31}$ values
versus input period from the simulations for the outer field. These
results show that there are no biases in our data resulting in
systematic errors or degeneracies in the fit $\phi_{31}$ values from
TFF. The simulations for the inner field conclude the same. Thus, we
expect no systematic errors in the metallicities calculated from $\phi_{31}$.

\begin{figure}[!ht]
\epsscale{0.95}
\plotone{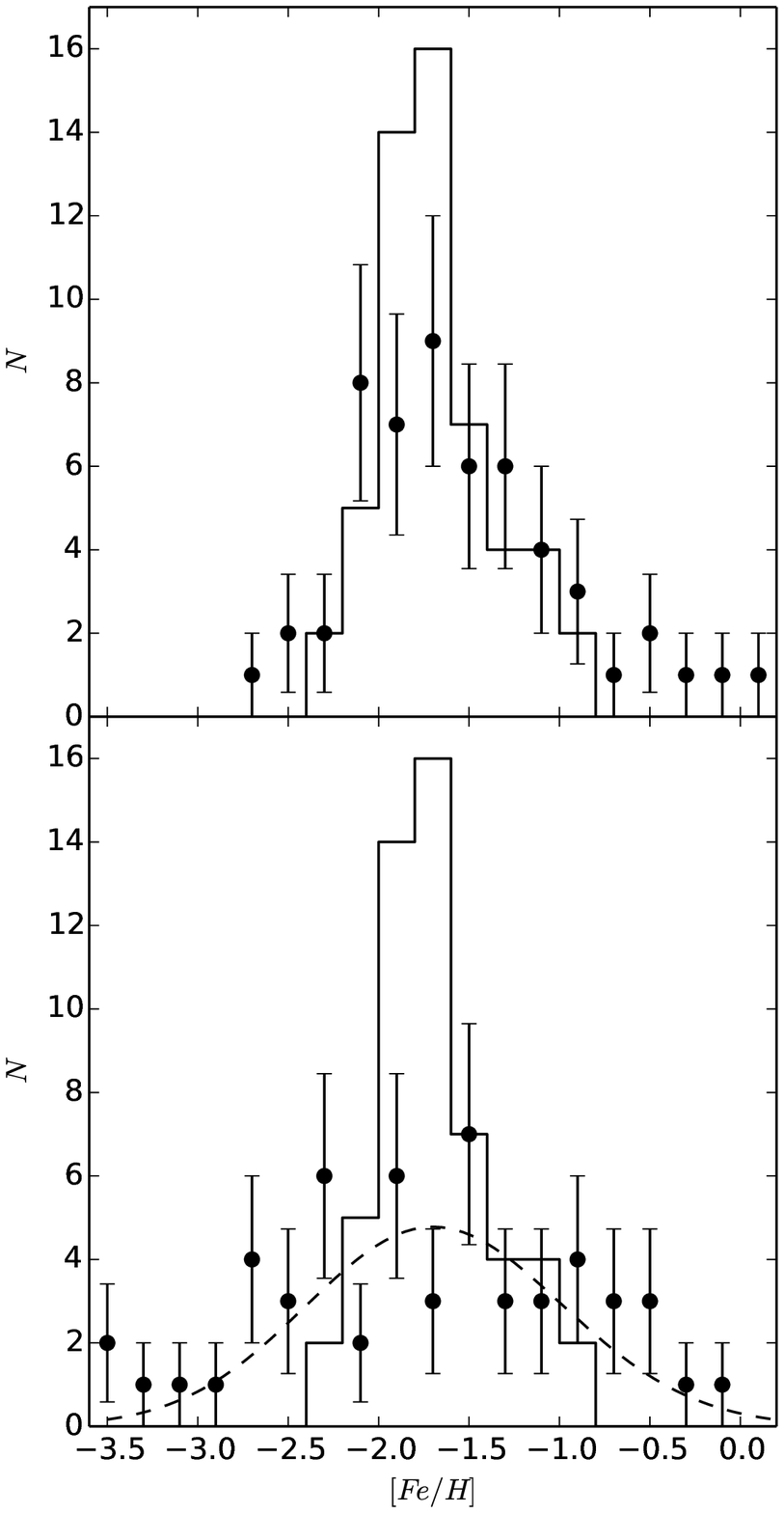}
\caption{{\small The MDF of the probable RRab stars in Phoenix. \emph{Top:}
  The MDF calculated using the \citet{alc00} relation (solid) using the FITLC
  periods and amplitudes compared with the MDF calculated
  using the \citet{jur96} relation (points) using the FITLC periods and
  $\phi_{31}$ determined by TFF. \emph{Bottom:} Same instead using the
  \citet{nem13} relation (points). The dashed curve shows the
  generalized histogram of the \citet{alc00} MDF using the error in
  $\langle [Fe/H]\rangle$ adopting the \citet{nem13} relation. The MDFs
  with error bars show the
  binned histograms with 1-$\sigma$ Poisson errors.}\label{fig:mdfcomp}}
\end{figure}

In order to compare the abundances from the different
relations, we show the MDF resulting from each calculation in Figure
\ref{fig:mdfcomp}. While the results from the RR Lyrae
candidates have greater signal (less Poisson noise), this set of
candidates is likely contaminated by various other variable types
(e.g. the likely anomalous Cepheid discussed in Section
\ref{sub:bailey}). Therefore, in an effort to compare abundances of
bona fide RR Lyrae stars only,  we exclusively refer to the probable RR
Lyrae results in this comparison as well as in the rest of this
paper. The MDFs in Figure \ref{fig:mdfcomp}
show binned histograms in each case using a bin size of 0.2 dex. 

Qualitatively, it is apparent that the \citet{jur96} MDF
resembles the overall shape of the \citet{alc00}. Meanwhile the
\citet{nem13} MDF has much more noise with no discernible shape
resembling the other two.  The mean metallicity of the RRab stars using the \citet{jur96}
relation is $\langle [Fe/H]\rangle = -1.55 \pm 0.03$ dex. Using
the \citet{nem13} relation, the mean metallicity becomes $\langle
[Fe/H]\rangle = -1.75 \pm 0.64$ dex (uncertainties represent errors in
period
and $\phi_{31}$ propagated through the metallicity
calculation summed with the standard error of the mean in
quadrature). In comparison the mean abundance
from the \citet{alc00} relation yields 
$\langle [Fe/H]\rangle = -1.68 \pm 0.06$ dex, which is significantly
closer to the other
abundance determinations previously mentioned ($[Fe/H] = -1.81 \pm 0.10$ dex and $[Fe/H] = -1.7
\pm 0.2$ dex; \citet{hel99,
hid09}) than the mean abundance using the \citet{jur96}
relation. \citet{jur96} point out that their relation predicts
systematically higher abundances for lower metallicity ($[Fe/H]
\lesssim$ -2.0 dex) RRab stars when
compared with the spectroscopic abundances, which could explain why
the metallicities calculated using it are higher than the other determinations.

On the other hand, the mean abundance using the \citet{nem13} relation
lies closer to other determinations. However, it has an
extremely large error, and its corresponding MDF appears significantly
more noisy than the other two. We suspect this to be resultant from the characteristics of the
dataset used in this study. Namely, while the data are high-quality in
the sense that the photometric errors are low, there are few data
points available for a Fourier decomposition to accurately determine
the shape of the light curves even with TFF. Especially in the case of
the \citet{nem13} relation where there is a quadratic $\phi_{31}$ term as well as a $P\phi_{31}$ term
, these errors become large and are probably
responsible for the poor quality of that corresponding MDF. 

In order
to test this speculation, we performed the following test. We constructed a generalized histogram for the
\citet{alc00} MDF assigning to each metallicity measurement an error
equal to the error in $\langle [Fe/H]\rangle$ using the \citet{nem13}
relation. This is shown as the dashed curve in Figure \ref{fig:mdfcomp}. That this generalized histogram agrees with the
binned histogram of the \citet{nem13} MDF supports the assertion that the two are
in agreement, and the errors in $\phi_{31}$ act to add more noise to
the MDF. Considering the results of this comparison, we consider the
MDF using the \citet{alc00} relation more accurate than the two using
$\phi_{31}$, and refer to it exclusively for the duration of this paper.

\section{Discussion}
\label{sec:disc}

A major goal of this work is to probe the early chemical evolution
of Phoenix through the RR Lyrae stars and compare their properties to
the analysis of \citet{hid09}. In particular, they perform a detailed
study of the SFH of Phoenix as a function of time and position in the
dwarf galaxy. In this section, we compare the chemical evolution law
(CEL) at early times ($\gtrsim$ 10 Gyr ago) derived from their
analysis with our results from the RRab stars.

First, we checked for any variation in RR Lyrae star properties with
galactocentric distance since Phoenix has been observed to contain a
stellar population gradient \citep{ort88, mart99, hel99}. Specifically,
younger stars are centrally concentrated while the older RGB stars are present
throughout the entire galaxy.
We further investigated this gradient for the very old stars by
examining the properties of the RR Lyrae stars as a function of 
galactocentric distance. We found no trends in
metallicity, pulsation period, or pulsation amplitude for the RR Lyrae
stars in Phoenix as a function of galactocentric distance, consistent
with recent work by \citet{hid13} which determined that the oldest
stellar populations in dwarf transition-type galaxies, specifically
Phoenix, are coeval at all galactocentric radii. This is also consistent with our analysis in
Section \ref{sub:bailey} where we found no significant difference in
RRab periods between the two observed fields. This implied spatial homogeneity of the
RRab population in Phoenix justifies combining the RRab stars from
both observed fields to perform the subsequent analysis of the CEL of Phoenix.

We note that in their analysis, \citet{hid09} constructed a grid of 16
different parameterizations of the synthetic CMDs used to calculate the
SFH of Phoenix. They average these solutions for the final solution
and take the 1-$\sigma$ dispersion in the solutions as the uncertainty associated with
their model SFH. Using these methods, they calculate the metallicity
of Phoenix at a look-back time of 10 Gyr to be $\langle [M/H]\rangle = -1.75 \pm
.11$ dex (with the assumption that $[Fe/H] = [M/H]$). They compare this CEL with several
chemical evolution models, of which only the closed-box and the outflow
models are compatible. This coupled with the overall shape of the RR
Lyrae MDF motivated our choice of the closed-box as our fiducial
model.

Since \citet{hid09} did not consider the effects of alpha-enhancement
in their models, we do the same for the sake of comparison and consider $[M/H] = [Fe/H]$ using the
metallicities, $[Fe/H]$, calculated for the RRab stars using Equation
\ref{eq:met} in Section \ref{sub:met}. We fit to the RRab MDF a closed-box
model of the following form:
\begin{equation} \label{eq:cb}
\frac{dN}{d[M/H]} \sim \frac{Z - Z_0}{p}e^{-(Z-Z_0)/p}
\end{equation}
where $Z_0$ is the initial metal abundance of the system, and $p$ is the
yield. The sum of these two quantities equals the average metal abundance
of the system so that $\langle Z\rangle = p + Z_0$
\citep{pag09}. Therefore, the peak of this abundance distribution in
logarithmic space is at $\langle [M/H]\rangle = log[(p +
Z_0)/Z_{\odot}]$. 

\begin{figure}[!ht]
\epsscale{1.1}
\plotone{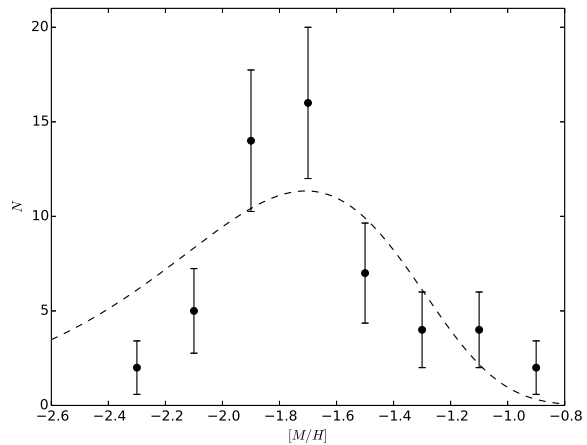}
\caption{{\small The MDF of the probable RRab stars in Phoenix. The MDF shows the
  binned histograms of the FITLC calculated metallicities with
  1-$\sigma$ Poisson error bars. The dashed
  line shows the best fit pure closed-box chemical evolution model to the
  MDF.}\label{fig:purecbmdf}}
\end{figure}

Figure \ref{fig:purecbmdf} shows a pure closed-box with no pre-enrichment,
where $Z_0 = 0$. A least-squares fit of this model to the data results
in a yield of $p = 0.00038$ ($\langle [M/H]\rangle = -1.71$  dex). This best
fit is plotted over the MDF in Figure \ref{fig:purecbmdf}. While the resultant value
for the mean metallicity does agree with the CEL from
\citet{hid09}, this model clearly fails to reproduce the shape of the RRab MDF. 
In particular, the pure
closed-box is marked by a significant low metallicity tail that is not
observed in the RRab MDF. This issue, the so-called G-dwarf problem
\citep{van62, sch63}, is known to be ubiquitous in galaxies
\citep{har00, bm98}, and has had extensive study devoted to its
resolution. Multiple solutions to this problem have been proposed, and
we proceed to investigate one such scenario for Phoenix, namely the
pre-enrichment scenario.

\begin{figure}[!ht]
\plotone{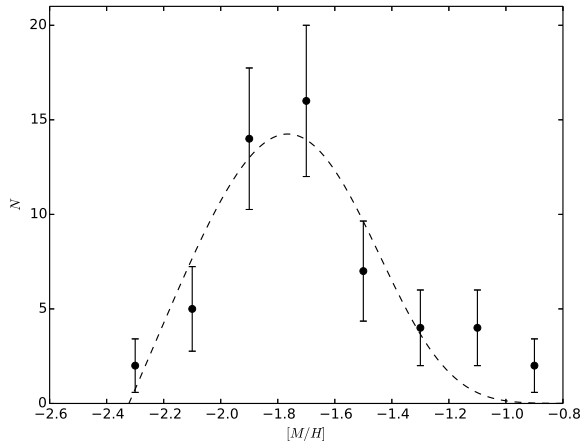}
\caption{The MDF of the probable RRab stars in Phoenix with a
  pre-enriched closed box model fit. This model clearly fits MDF
  better especially capturing the low metallicity tail better than the
  pure closed-box model.\label{fig:cbmdf}}
\end{figure}

This analysis indicates that the MDF of Phoenix at an early age is
not consistent with a pure closed-box CEL. To attempt to reconcile the
MDF with a chemical evolution model, we consider the scenario in which
the galaxy was born with a non-zero metallicity, $Z_0 \neq 0$. Under
this assumption, we
again fit a closed-box given by Equation \ref{eq:cb} to the MDF of
Phoenix, but this time allowing $Z_0$ to
vary as a free parameter. We find that a least-squares fit of a
pre-enriched closed-box model agrees well with the RRab MDF of
Phoenix as illustrated in Figure \ref{fig:cbmdf}. The probable RRab MDF is best fit with a closed-box with
an initial metal abundance of $Z_0 = 0.00001 \pm 0.00001$ ($[M/H] =
-2.32 \pm 0.06$ dex) and a yield of $p =
0.00024 \pm 0.00003$, which translates to a mean metallicity of
$\langle [M/H]\rangle = -1.77
\pm 0.05$ dex 
for the system. The uncertainties quoted for these metal abundances represent
the standard deviations calculated from the covariance matrix of the
model parameters of the fit ($\sqrt{\sigma^2}$), while that of the
metallicity represents their propagation through the metallicity calculation. This
agrees very well with the CEL of Phoenix at early times from \citet{hid09}
yielding a metallicity of $\langle [M/H]\rangle = -1.79 \pm
0.12$ dex. This value was obtained by averaging the metallicities of their
model for ages $10-13$ Gyr.

Our analysis thus indicates that the old epoch of star formation
associated with the RR Lyrae variables present in Phoenix is
consistent with a pre-enriched closed-box model. We now consider the
physical interpretation of this scenario in the context of the current
understanding of galaxy formation and evolution. The over estimation
of low metallicity stars by a pure closed-box model, the ubiquitous
G-dwarf problem, is reconciled with the observed RRab MDF if some form of chemical
pre-enrichment had occurred in Phoenix prior to the formation of
these old stars. 

Many studies have previously investigated possible
scenarios leading to such pre-enrichment in galaxies. \citet{tru71}
first proposed the idea of a generation of pre-galactic star formation
which preferentially formed massive stars. This early generation of
high mass stars would then evolve
rapidly and promptly enrich the early galaxy forming
environments. Thus such galaxies would form from material previously
enriched by this initial, short lived generation of stars. In the
case of Phoenix, such prompt initial enrichment would manifest itself as an
RR Lyrae MDF described by a pre-enriched chemical evolution
model. Another possible source of chemical pre-enrichment could be
extra-galactic. In particular, one might imagine a scenario in
which chemically enriched material from a further evolved, neighboring
galaxy is transferred to the system . However, we point out that the
isolated status of Phoenix in the Local Group makes this situation
unlikely.

Our analysis concludes that Phoenix chemically evolved similar to a pre-enriched
closed-box as evident from the RR Lyrae MDF. The parameters of the
best fit closed-box model to this MDF agree with the analysis of
\citet{hid09}, supporting their SFH analysis for the early evolution
of Phoenix. We interpret this result as evidence that the prompt
initial enrichment scenario for galaxies first proposed by \citet{tru71} likely
occurred in Phoenix.

\section{Conclusions}
\label{sec:conc}

We have presented the first detailed study of the RR Lyrae stars
populating the Phoenix dwarf galaxy. We used light curve template
fitting routines to identify and characterize RR Lyrae variables
within Phoenix using archival WFPC2 data. The cadence and phase
coverage of these data coupled with our fitting routines allowed us to
increase the number of highly probable RR Lyrae stars to $78$, with
$\sim 40$ more RR Lyrae candidates observed outside of the instability
strip. The mean periods calculated for the two types of RR Lyrae found
in Phoenix are $\langle P_{ab}\rangle = 0.595 \pm 0.032$ d and
$\langle P_{ab}\rangle = 0.353 \pm 0.002$ d for the ab- and c-types,
respectively. 

We have used the properties of the RR Lyrae population within Phoenix
to probe its behavior at early times. Using the best fit light curve
properties, we have constructed the Bailey
diagram for the RR Lyrae stars in Phoenix which displays the RRab
stars apparently following the OoI relation. However, the position of
Phoenix in the [Fe/H]-$\langle P_{ab}\rangle$ plane lies within the Oosterhoff gap,
indicating that it is likely of intermediate Oosterhoff type. This is
consistent with most other Milky Way dwarf satellite galaxies, however
the discrepancy between the Bailey diagram and the calculated values
of $[Fe/H]$ and $\langle P_{ab}\rangle$ warrants further investigation.

We used the minimum light colors of the RRab stars in Phoenix to
estimate the line of sight reddening to the galaxy. Using this method,
we calculated the reddening to Phoenix to be $E(V-I) = 0.07 \pm
0.06$ mag. This does not agree well with the
\citet{sch98} maps. However, this does qualitatively agree with previous
determinations of the reddening to Phoenix indicating internal sources
of extinction resulting in such a discrepancy.

We also studied the RR Lyrae star properties as a function
of galactocentric radius in Phoenix and found no significant trends,
consistent with the previously observed stellar population gradient
for old stars in the dwarf galaxy. In particular, we found no
significant trends in $\langle P_{ab}\rangle$, [Fe/H], or $A_V$ with
respect to distance from the center of Phoenix in our RR Lyrae
sample. We did however find a small but significant
difference in mean RRc period, $\langle P_c\rangle$, between our two
observed fields. 
 
Finally, we fit a closed-box chemical evolution model to the MDF of
the RRab stars in Phoenix. Using the pulsation periods and amplitudes of the RRab stars, we
calculated their metallicities using the period-amplitude-metallicity
relation from \citet{alc00}. We obtained a mean metallicity for the
probable RRab stars in Phoenix of $\langle [Fe/H]\rangle = -1.68 \pm
0.06$ dex. We found that a pure closed-box devoid of
pre-enrichment poorly fit the MDF. However, a pre-enriched closed-box
fits the MDF much better. The average metallicity associated
with this best fit pre-enriched model, $[M/H] = -1.77 \pm 0.05$ dex, agrees well with the
CEL for Phoenix derived by \citet{hid09}, supporting  the notion that
this galaxy chemically evolved similar to a pre-enriched, closed-box at a
young age. Due to the isolated nature of Phoenix, we speculate that
this pre-enrichment was not likely from any external source. Instead,
we suggest that the prompt initial enrichment scenario in which
material that contributed to the formation of Phoenix likely experienced an early generation of star
formation marked by preferentially massive stars. 

\acknowledgements

We are grateful to Bart Pritzl and Sebastian Hidalgo for providing 
constructive comments that greatly improved an earlier version of this
manuscript. We also thank G\'{e}za Kov\'{a}cs for his TFF software and
helping us by modifying it to ensure accurate fitting on our dataset. 

This work was supported by KASI-Carnegie Fellowship Program jointly managed by Korea
Astronomy and Space Science Institute (KASI) and the Observatories of the Carnegie Institution for Science.

\begin{deluxetable}{cccc}
\tabletypesize{\scriptsize}
\tablecaption{An example time-series photometry set for one probable
  (ID: V01852) RR Lyrae star in Phoenix. The full set of these
  for all RR Lyrae star candidates is included in the online
  version of the journal.\label{tbl:timephot}}
\tablewidth{0pt}
\tablehead{
\colhead{Filter} & \colhead{HJD - 2451900 (days)} & \colhead{Star
  magnitude} & \colhead{Magnitude error}
}
\startdata
F555W & 1925.66641 & 24.019 & 0.235\\ 
F555W & 1925.67428 & 23.707 & 0.03\\ 
F555W & 1925.69094 & 23.2 & 0.022\\ 
F555W & 1925.73805 & 23.164 & 0.043\\ 
F555W & 1925.7561 & 23.291 & 0.022\\ 
F555W & 1925.8054 & 23.523 & 0.026\\ 
F555W & 1925.82346 & 23.598 & 0.026\\ 
F555W & 1925.87276 & 23.687 & 0.028\\ 
F555W & 1925.89082 & 23.495 & 0.168\\ 
F555W & 1925.93943 & 23.921 & 0.033\\ 
F555W & 1925.95748 & 24.021 & 0.034\\ 
F814W & 1926.62562 & 23.465 & 0.047\\ 
F814W & 1926.67631 & 23.555 & 0.052\\ 
F814W & 1926.69356 & 23.548 & 0.047\\ 
F814W & 1926.74425 & 23.55 & 0.05\\ 
F814W & 1926.7623 & 23.441 & 0.044\\ 
F814W & 1926.81161 & 22.85 & 0.031\\ 
F814W & 1926.82966 & 22.921 & 0.031\\ 
F814W & 1926.87827 & 22.99 & 0.167\\ 
F814W & 1926.89632 & 23.161 & 0.036\\ 
F814W & 1926.94563 & 23.231 & 0.039\\ 
F814W & 1926.96368 & 23.222 & 0.038\\ 
F814W & 1927.01229 & 23.358 & 0.043\\ 
\enddata
\end{deluxetable}

\begin{deluxetable}{cllcccccc}
\tabletypesize{\scriptsize}
\tablecaption{IDs, positions, and light curve parameters for the 
probable RR Lyrae stars within Phoenix.\label{tbl:rrlconf}}
\tablewidth{0pt}
\tablehead{
\colhead{ID} & \colhead{RA (J2000)} &
\colhead{Dec (J2000)} & \colhead{Type} & \colhead{Period (d)} 
& \colhead{$A_V$ (mag)} & \colhead{$A_I$ (mag)} &
\colhead{$\langle V\rangle$ (mag)}
& \colhead{$\langle I\rangle$ (mag)}
}
\startdata
V00030 & 1 51 10.3023 & -44 23 46.811 & ab & 0.55 & 0.719 & 0.3 & 23.957 & 23.444\\ 
V01828 & 1 51 9.385 & -44 24 7.623 & ab & 0.577 & 1.145 & 0.716 & 23.747 & 23.242\\ 
V01850 & 1 51 12.7286 & -44 24 57.335 & c & 0.383 & 0.525 & 0.328 & 23.636 & 23.204\\ 
V01851 & 1 51 7.3409 & -44 24 45.572 & ab & 0.657 & 0.642 & 0.379 & 23.767 & 23.153\\ 
V01852 & 1 51 10.1339 & -44 24 39.903 & ab & 0.55 & 1.167 & 0.729 & 23.728 & 23.307\\ 
V01862 & 1 51 10.6558 & -44 25 27.132 & ab & 0.6 & 0.877 & 0.548 & 23.826 & 23.246\\ 
V01867 & 1 51 10.1103 & -44 24 23.18 & ab & 0.642 & 0.728 & 0.455 & 23.684 & 23.101\\ 
V01868 & 1 51 9.5417 & -44 25 12.712 & ab & 0.571 & 0.89 & 0.556 & 23.783 & 23.27\\ 
V01870 & 1 51 7.0235 & -44 25 7.425 & ab & 0.591 & 0.819 & 0.512 & 23.779 & 23.264\\ 
V01877 & 1 51 11.3796 & -44 25 1.509 & c & 0.356 & 0.509 & 0.318 & 23.698 & 23.292\\ 
V01889 & 1 51 11.3589 & -44 24 40.242 & c & 0.36 & 0.492 & 0.307 & 23.735 & 23.273\\ 
V01900 & 1 51 9.4811 & -44 24 45.698 & c & 0.34 & 0.473 & 0.349 & 23.72 & 23.337\\ 
V01904 & 1 51 6.6992 & -44 25 10.371 & ab & 0.591 & 0.746 & 0.466 & 23.713 & 23.178\\ 
V01906 & 1 51 12.6181 & -44 25 15.33 & c & 0.37 & 0.626 & 0.31 & 23.735 & 23.251\\ 
V01909 & 1 51 8.8591 & -44 25 23.986 & ab & 0.688 & 0.556 & 0.348 & 23.87 & 23.233\\ 
V01926 & 1 51 11.6177 & -44 24 39.268 & ab & 0.575 & 0.375 & 0.235 & 23.806 & 23.21\\ 
V01933 & 1 51 10.2845 & -44 24 58.76 & c & 0.364 & 0.35 & 0.131 & 23.785 & 23.463\\ 
V08484 & 1 51 5.372 & -44 24 8.984 & ab & 0.528 & 1.104 & 0.69 & 23.791 & 23.334\\ 
V08486 & 1 51 8.4435 & -44 24 16.779 & ab & 0.598 & 0.989 & 0.618 & 23.805 & 23.257\\ 
V08503 & 1 51 1.9574 & -44 24 37.401 & ab & 0.584 & 0.896 & 0.56 & 23.814 & 23.298\\ 
V08517 & 1 51 6.4885 & -44 24 49.23 & ab & 0.575 & 0.52 & 0.325 & 23.792 & 23.194\\ 
V08519 & 1 51 3.5394 & -44 23 54.399 & ab & 0.54 & 1.194 & 0.746 & 23.743 & 23.297\\ 
V08525 & 1 51 7.9508 & -44 24 9.073 & c & 0.33 & 0.521 & 0.326 & 23.742 & 23.378\\ 
V08527 & 1 51 4.2859 & -44 24 20.434 & c & 0.29 & 0.258 & 0.176 & 23.739 & 23.391\\ 
V08535 & 1 51 3.9698 & -44 24 26.839 & ab & 0.55 & 0.6 & 0.351 & 23.911 & 23.389\\ 
V08547 & 1 51 6.6909 & -44 24 13.971 & ab & 0.52 & 0.705 & 0.42 & 23.836 & 23.297\\ 
V08548 & 1 51 7.4302 & -44 24 33.553 & ab & 0.527 & 0.55 & 0.343 & 23.873 & 23.365\\ 
V13413 & 1 51 5.4186 & -44 23 26.427 & ab & 0.55 & 0.9 & 0.5 & 23.853 & 23.411\\ 
V13419 & 1 51 8.8069 & -44 23 35.074 & ab & 0.604 & 0.761 & 0.475 & 23.788 & 23.163\\ 
V13426 & 1 51 8.3281 & -44 23 36.738 & c & 0.367 & 0.435 & 0.272 & 23.785 & 23.311\\ 
V13428 & 1 51 5.9148 & -44 23 32.945 & ab & 0.633 & 0.452 & 0.282 & 23.878 & 23.248\\ 
V13439 & 1 51 9.8959 & -44 22 48.219 & c & 0.27 & 0.283 & 0.166 & 23.795 & 23.421\\ 
V00182 & 1 51 10.9012 & -44 26 16.739 & c & 0.352 & 0.43 & 0.269 & 23.756 & 23.305\\ 
V00220 & 1 51 8.3127 & -44 26 19.618 & ab & 0.711 & 0.738 & 0.461 & 24.0 & 23.471\\ 
V00221 & 1 51 8.4787 & -44 26 30.691 & ab & 0.64 & 0.6 & 0.375 & 23.994 & 23.542\\ 
V00230 & 1 51 8.9078 & -44 26 37.951 & ab & 0.565 & 0.908 & 0.556 & 23.785 & 23.346\\ 
V04487 & 1 51 7.5186 & -44 27 6.035 & ab & 0.56 & 1.2 & 0.752 & 23.27 & 22.728\\ 
V04565 & 1 51 9.8315 & -44 27 33.477 & ab & 0.716 & 0.752 & 0.47 & 23.598 & 23.067\\ 
V04566 & 1 51 7.132 & -44 27 39.686 & c & 0.393 & 0.482 & 0.301 & 23.609 & 23.197\\ 
V04629 & 1 51 6.1442 & -44 27 19.921 & c & 0.341 & 0.494 & 0.309 & 23.745 & 23.328\\ 
V04641 & 1 51 10.1578 & -44 27 25.703 & ab & 0.591 & 0.834 & 0.55 & 23.682 & 23.193\\ 
V04642 & 1 51 11.5694 & -44 27 16.29 & c & 0.38 & 0.436 & 0.241 & 23.67 & 23.27\\ 
V04663 & 1 51 8.3945 & -44 27 5.552 & ab & 0.557 & 0.637 & 0.398 & 23.708 & 23.112\\ 
V04664 & 1 51 6.8452 & -44 27 25.634 & c & 0.36 & 0.418 & 0.23 & 23.769 & 23.338\\ 
V04666 & 1 51 7.0685 & -44 26 54.726 & ab & 0.615 & 0.684 & 0.428 & 23.745 & 23.209\\ 
V04675 & 1 51 5.7521 & -44 27 21.587 & c & 0.4 & 0.355 & 0.19 & 23.74 & 23.3\\ 
V04677 & 1 51 11.3829 & -44 27 33.179 & c & 0.33 & 0.4 & 0.244 & 23.75 & 23.294\\ 
V04694 & 1 51 9.5196 & -44 27 20.383 & ab & 0.57 & 1.237 & 0.776 & 23.469 & 23.035\\ 
V04702 & 1 51 6.9907 & -44 27 10.772 & ab & 0.65 & 0.425 & 0.266 & 23.818 & 23.267\\ 
V04708 & 1 51 8.0198 & -44 27 19.846 & ab & 0.64 & 0.679 & 0.4 & 23.741 & 23.171\\ 
V04711 & 1 51 6.2346 & -44 27 44.12 & ab & 0.64 & 0.681 & 0.405 & 23.691 & 23.159\\ 
V05076 & 1 51 7.2067 & -44 27 44.776 & ab & 0.662 & 0.75 & 0.45 & 24.118 & 23.827\\ 
V15025 & 1 51 4.5119 & -44 27 24.002 & c & 0.4 & 0.409 & 0.223 & 23.419 & 23.025\\ 
V15061 & 1 51 3.5596 & -44 26 36.829 & ab & 0.524 & 0.939 & 0.587 & 23.684 & 22.989\\ 
V15146 & 1 51 2.0041 & -44 27 22.887 & c & 0.259 & 0.425 & 0.266 & 23.689 & 23.089\\ 
V15193 & 1 51 3.5748 & -44 27 7.108 & ab & 0.6 & 0.595 & 0.372 & 23.769 & 23.235\\ 
V15217 & 1 51 3.8976 & -44 26 53.091 & c & 0.35 & 0.43 & 0.269 & 23.702 & 23.291\\ 
V15222 & 1 51 0.9742 & -44 26 22.131 & c & 0.307 & 0.346 & 0.216 & 23.723 & 23.29\\ 
V15239 & 1 51 4.6409 & -44 26 44.189 & ab & 0.6 & 0.492 & 0.279 & 23.737 & 23.225\\ 
V15264 & 1 51 1.9697 & -44 26 52.379 & ab & 0.6 & 0.959 & 0.599 & 23.799 & 23.361\\ 
V15265 & 1 51 4.3707 & -44 27 40.149 & ab & 0.571 & 1.166 & 0.729 & 23.592 & 23.017\\ 
V15267 & 1 51 2.8094 & -44 27 35.197 & ab & 0.616 & 0.705 & 0.44 & 23.835 & 23.333\\ 
V15297 & 1 51 2.8737 & -44 26 41.092 & ab & 0.65 & 0.898 & 0.55 & 23.64 & 23.146\\ 
V15330 & 1 51 1.1669 & -44 26 23.008 & ab & 0.53 & 1.058 & 0.657 & 23.73 & 23.18\\ 
V15334 & 1 51 1.3606 & -44 26 41.063 & ab & 0.581 & 1.028 & 0.643 & 23.796 & 23.286\\ 
V15397 & 1 50 59.5913 & -44 27 8.637 & ab & 0.56 & 1.063 & 0.665 & 23.746 & 23.293\\ 
V24705 & 1 51 8.7395 & -44 25 37.304 & ab & 0.72 & 0.348 & 0.218 & 23.566 & 23.022\\ 
V24730 & 1 51 7.7841 & -44 25 59.551 & ab & 0.61 & 0.999 & 0.624 & 23.707 & 23.271\\ 
V24740 & 1 51 7.8297 & -44 25 48.525 & ab & 0.56 & 0.884 & 0.54 & 23.77 & 23.142\\ 
V24758 & 1 51 7.6053 & -44 25 57.245 & ab & 0.558 & 1.161 & 0.725 & 23.48 & 22.962\\ 
V24796 & 1 51 6.4818 & -44 26 7.381 & c & 0.4 & 0.466 & 0.244 & 23.693 & 23.277\\ 
V24816 & 1 51 7.442 & -44 26 28.76 & ab & 0.574 & 0.857 & 0.535 & 23.755 & 23.15\\ 
V24819 & 1 51 5.361 & -44 25 31.527 & ab & 0.592 & 0.857 & 0.535 & 23.761 & 23.281\\ 
V24830 & 1 51 3.9655 & -44 26 15.488 & ab & 0.55 & 1.242 & 0.744 & 23.726 & 23.285\\ 
V24859 & 1 51 4.7892 & -44 26 19.749 & ab & 0.56 & 1.156 & 0.722 & 23.826 & 23.299\\ 
V24860 & 1 51 5.4346 & -44 25 35.49 & c & 0.397 & 0.48 & 0.3 & 23.73 & 23.331\\ 
V24867 & 1 51 1.9257 & -44 25 54.047 & ab & 0.631 & 0.419 & 0.262 & 23.874 & 23.224\\ 
V24930 & 1 51 5.4548 & -44 25 43.291 & c & 0.36 & 0.387 & 0.242 & 23.802 & 23.355\\ 
\enddata
\end{deluxetable}

\begin{deluxetable}{cllcccccc}
\tabletypesize{\scriptsize}
\tablecaption{Same as Table \ref{tbl:rrlconf} for the remaining RR Lyrae candidates.\label{tbl:rrlcan}}
\tablewidth{0pt}
\tablehead{
\colhead{ID} & \colhead{RA (J2000)} &
\colhead{Dec (J2000)} & \colhead{Type} & \colhead{Period (d)} 
& \colhead{$A_V$ (mag)} & \colhead{$A_I$ (mag)} &
\colhead{$\langle V\rangle$ (mag)}
& \colhead{$\langle I\rangle$ (mag)}
}
\startdata
V00026 & 1 51 12.2262 & -44 23 50.628 & ab & 0.55 & 0.925 & 0.492 & 23.91 & 23.219\\ 
V00029 & 1 51 10.6934 & -44 24 2.009 & ab & 0.566 & 0.806 & 0.504 & 23.903 & 23.2\\ 
V01737 & 1 51 7.3919 & -44 25 9.917 & ab & 0.81 & 0.7 & 0.426 & 23.009 & 22.375\\ 
V01920 & 1 51 8.6178 & -44 25 7.858 & ab & 0.606 & 1.071 & 0.669 & 23.707 & 23.245\\ 
V01922 & 1 51 14.8559 & -44 24 36.817 & ab & 0.54 & 0.94 & 0.593 & 23.787 & 23.247\\ 
V01935 & 1 51 12.8491 & -44 25 2.949 & ab & 0.53 & 0.988 & 0.594 & 23.799 & 23.253\\ 
V01936 & 1 51 10.6045 & -44 25 33.379 & ab & 0.636 & 0.823 & 0.514 & 23.786 & 23.268\\ 
V01964 & 1 51 9.6648 & -44 24 33.275 & ab & 0.54 & 1.17 & 0.727 & 23.772 & 23.287\\ 
V01987 & 1 51 10.7769 & -44 25 12.145 & ab & 0.77 & 0.288 & 0.128 & 24.213 & 23.325\\ 
V08460 & 1 51 6.265 & -44 24 1.739 & ab & 0.67 & 0.922 & 0.946 & 23.68 & 23.014\\ 
V08478 & 1 51 5.1497 & -44 24 16.041 & ab & 0.501 & 1.285 & 0.803 & 23.83 & 23.338\\ 
V08499 & 1 51 4.6759 & -44 24 43.511 & ab & 0.67 & 0.597 & 0.373 & 23.721 & 23.152\\ 
V08518 & 1 51 6.2679 & -44 24 16.05 & ab & 0.648 & 0.7 & 0.231 & 23.733 & 23.202\\ 
V08553 & 1 51 6.2555 & -44 24 57.4 & ab & 0.65 & 0.955 & 0.317 & 23.739 & 23.257\\ 
V08557 & 1 51 5.5317 & -44 24 48.54 & ab & 0.71 & 0.626 & 0.391 & 23.807 & 23.305\\ 
V08558 & 1 51 3.2475 & -44 24 25.363 & ab & 0.55 & 0.833 & 0.51 & 23.765 & 23.232\\ 
V13436 & 1 51 9.5384 & -44 22 49.036 & ab & 0.555 & 1.0 & 0.618 & 23.806 & 23.194\\ 
V04221 & 1 51 10.9832 & -44 27 48.207 & ab & 0.629 & 0.675 & 0.4 & 23.123 & 22.181\\ 
V04269 & 1 51 6.6831 & -44 27 14.035 & c & 0.646 & 0.16 & 0.1 & 23.322 & 22.453\\ 
V04354 & 1 51 12.5259 & -44 27 27.815 & ab & 0.641 & 0.572 & 0.358 & 23.375 & 22.452\\ 
V04406 & 1 51 9.084 & -44 26 49.047 & ab & 0.665 & 0.552 & 0.345 & 23.364 & 22.599\\ 
V04417 & 1 51 5.9308 & -44 27 32.915 & ab & 0.661 & 0.554 & 0.31 & 23.425 & 22.577\\ 
V04418 & 1 51 7.4437 & -44 27 42.772 & ab & 0.506 & 1.06 & 0.662 & 23.63 & 23.071\\ 
V04507 & 1 51 11.0036 & -44 27 49.6 & ab & 0.55 & 0.16 & 0.1 & 23.652 & 22.78\\ 
V04570 & 1 51 7.5821 & -44 27 9.353 & ab & 0.603 & 0.733 & 0.458 & 23.755 & 23.177\\ 
V04571 & 1 51 7.5345 & -44 27 41.459 & ab & 0.67 & 0.294 & 0.184 & 23.631 & 22.95\\ 
V04712 & 1 51 9.3698 & -44 27 48.068 & c & 0.334 & 0.16 & 0.1 & 24.027 & 23.101\\ 
V04721 & 1 51 7.9728 & -44 27 6.884 & ab & 0.614 & 0.628 & 0.393 & 23.741 & 23.204\\ 
V04741 & 1 51 9.3652 & -44 27 36.423 & ab & 0.665 & 0.496 & 0.31 & 23.91 & 23.023\\ 
V04799 & 1 51 11.1361 & -44 27 36.472 & ab & 0.481 & 1.379 & 0.871 & 23.825 & 23.342\\ 
V14821 & 1 51 4.0875 & -44 27 17.392 & ab & 0.71 & 0.391 & 0.244 & 23.165 & 22.388\\ 
V14932 & 1 51 1.6833 & -44 26 10.717 & ab & 0.79 & 0.3 & 0.15 & 23.602 & 22.542\\ 
V15082 & 1 51 5.0753 & -44 26 52.258 & ab & 0.481 & 1.279 & 0.799 & 23.465 & 22.801\\ 
V15143 & 1 51 1.7939 & -44 26 53.144 & ab & 0.52 & 1.366 & 0.854 & 23.706 & 23.291\\ 
V15170 & 1 51 2.8397 & -44 26 38.914 & ab & 0.613 & 0.561 & 0.351 & 23.728 & 23.156\\ 
V15191 & 1 51 5.4653 & -44 26 36.46 & ab & 0.601 & 0.738 & 0.461 & 23.78 & 23.262\\ 
V15207 & 1 51 4.0704 & -44 27 40.983 & ab & 0.57 & 0.536 & 0.335 & 23.648 & 23.034\\ 
V15242 & 1 51 3.9535 & -44 26 58.767 & ab & 0.58 & 0.668 & 0.396 & 23.818 & 23.223\\ 
V15271 & 1 51 5.6198 & -44 26 51.548 & ab & 0.526 & 0.886 & 0.554 & 23.745 & 23.299\\ 
V24727 & 1 51 4.6408 & -44 25 29.496 & ab & 0.709 & 0.462 & 0.288 & 23.632 & 22.681\\ 
V24794 & 1 51 7.3438 & -44 26 23.126 & ab & 0.601 & 0.847 & 0.529 & 23.778 & 23.266\\ 
V24906 & 1 51 8.3403 & -44 26 0.238 & ab & 0.58 & 0.987 & 0.617 & 23.85 & 23.272\\ 
V24962 & 1 51 7.4637 & -44 26 26.23 & ab & 0.599 & 1.02 & 0.631 & 23.874 & 23.322\\ 
\enddata
\end{deluxetable}

\end{document}